\documentclass[prd,twocolumn,superscriptaddress,nofootinbib,reprint]{revtex4-1}

\usepackage{blindtext}

\usepackage{amsmath}

\usepackage{xspace}
\usepackage{xcolor}

\usepackage{todonotes}
\usepackage{aas_macros}

\usepackage[range-units=single,range-phrase= -- ,load-configurations=abbreviations,detect-weight=true,detect-family=true]{siunitx}

\usepackage{amsmath}
\usepackage{mathtools}
\newcommand{\la}{\ensuremath{\left\langle}}
\newcommand{\ra}{\ensuremath{\right\rangle}}

\newcommand{\lv}{\ensuremath{\left\lvert}}
\newcommand{\rv}{\ensuremath{\right\rvert}}

\newcommand{\ls}{\ensuremath{\left[}} 
\newcommand{\rs}{\ensuremath{\right]}}

\newcommand{\lp}{\ensuremath{\left(}} 
\newcommand{\rp}{\ensuremath{\right)}}

\renewcommand{\vec}[1]{\boldsymbol{#1}}
\newcommand{\mat}[1]{\boldsymbol{\mathbf{#1}}}

\newcommand{\hconj}{\ensuremath{\dagger}}

\newcommand{\vx}{\vec{x}}
\newcommand{\vs}{\vec{s}}
\newcommand{\vn}{\vec{n}}
\newcommand{\va}{\vec{a}}
\newcommand{\vv}{\vec{v}}
\newcommand{\vw}{\vec{w}}
\newcommand{\vk}{\vec{k}}

\newcommand{\mS}{\mat{S}}
\newcommand{\mN}{\mat{N}}
\newcommand{\mR}{\mat{R}}
\newcommand{\mC}{\mat{C}}
\newcommand{\mB}{\mat{B}}
\newcommand{\mU}{\mat{U}}
\newcommand{\mV}{\mat{V}}
\newcommand{\mE}{\mat{E}}
\newcommand{\mF}{\mat{F}}
\newcommand{\mJ}{\mat{J}}
\newcommand{\mI}{\mat{I}}

\newcommand{\mNh}{\mat{N}^{\scriptscriptstyle -\frac{1}{2}}}

\newcommand{\mQ}{\mat{Q}}
\newcommand{\mP}{\mat{P}}

\newcommand{\mCt}{\tilde{\mat{C}}}
\newcommand{\mSt}{\tilde{\mat{S}}}
\newcommand{\mNt}{\tilde{\mat{N}}}

\newcommand{\msigma}{\mat{\Sigma}}
\newcommand{\mLambda}{\mat{\Lambda}}
\newcommand{\mLambdat}{\tilde{\mat{\Lambda}}}

\newcommand{\vnhat}{\hat{\vec{n}}}
\newcommand{\vuhat}{\hat{\vec{u}}}

\newcommand{\vyhat}{\hat{\vec{y}}}
\newcommand{\vxhat}{\hat{\vec{x}}}

\newcommand{\vdhat}{\hat{\vec{d}}}

\newcommand{\vu}{\vec{u}}

\newcommand{\brsc}[1]{{\ensuremath{\scriptscriptstyle (#1)}}}

\newcommand{\calP}{\mathcal{P}}

\newcommand{\vnt}{\tilde{\vec{n}}}
\newcommand{\vvt}{\tilde{\vec{v}}}
\newcommand{\vxt}{\tilde{\vec{x}}}

\newcommand{\mBb}{\bar{\mat{B}}}
\newcommand{\vnb}{\bar{\vec{n}}}
\newcommand{\vvb}{\bar{\vec{v}}}

\newcommand{\mCb}{\bar{\mat{C}}}
\newcommand{\mNb}{\bar{\mat{N}}}

\newcommand{\mSb}{\bar{\mat{S}}}
\newcommand{\mFb}{\bar{\mat{F}}}

\DeclareMathOperator{\sinc}{sinc}
\DeclareMathOperator{\tri}{tri}

\DeclareMathOperator{\Tr}{Tr}

\DeclareMathOperator{\Cov}{Cov}

\newcommand{\vEps}{\vec{\varepsilon}}
\newcommand{\eps}{\varepsilon}

\newcommand{\vE}{\vec{E}}
\newcommand{\vH}{\vec{H}}
\newcommand{\vS}{\vec{S}}
\newcommand{\vA}{\vec{A}}

\newcommand{\dhn}{d^2\hat{n}}

\newcommand{\ihMpc}{\;h\:\mathrm{Mpc}^{-1}}

\usepackage{hyperref}
\usepackage[capitalise,noabbrev,nameinlink]{cleveref}

\renewcommand{\eqref}[1]{\cref{#1}}

\newcommand{\tcm}{\SI{21}{\centi\metre}\xspace}
\newcommand{\tcmt}{21$\,$cm\xspace}
\DeclareSIUnit\jansky{Jy}
\newcommand{\KLfull}{Karhunen-Lo\`{e}ve\xspace}

\begin{document}

\title{Coaxing Cosmic \tcmt Fluctuations from the Polarized Sky using $m$-mode Analysis}
  

\author{J. Richard Shaw}
\email{jrs65@cita.utoronto.ca} 
\affiliation{Canadian Institute for Theoretical Astrophysics, 60 St. George St., Toronto, ON M5S 3H8, Canada}
\author{Kris Sigurdson}
\affiliation{Department of Physics and Astronomy, University of British Columbia, Vancouver, BC V6T 1Z1, Canada}
\author{Michael Sitwell}
\affiliation{Department of Physics and Astronomy, University of British Columbia, Vancouver, BC V6T 1Z1, Canada}
\author{Albert Stebbins}
\affiliation{Theoretical Astrophysics Group, Fermi National Accelerator Laboratory, Batavia, IL 60510, USA}
\author{Ue-Li Pen}
\affiliation{Canadian Institute for Theoretical Astrophysics, 60 St. George St., Toronto, ON M5S 3H8, Canada}

\begin{abstract}

In this paper we continue to develop the $m$-mode formalism, a technique for
efficient and optimal analysis of wide-field transit radio telescopes,
targeted at \tcm cosmology. We extend this formalism to give an accurate
treatment of the polarised sky, fully accounting for the effects of
polarisation leakage and cross-polarisation. We use the geometry of the
measured set of visibilities to project down to pure temperature modes on the
sky, serving as a significant compression, and an effective first filter of
polarised contaminants. As in our previous work, we use the $m$-mode formalism
with the \KLfull transform to give a highly efficient method for foreground
cleaning, and demonstrate its success in cleaning realistic polarised skies
observed with an instrument suffering from substantial off axis polarisation
leakage. We develop an optimal quadratic estimator in the $m$-mode formalism,
which can be efficiently calculated using a Monte-Carlo technique. This is
used to assess the implications of foreground removal for power spectrum
constraints where we find that our method can clean foregrounds well below the
foreground wedge, rendering only scales $k_\parallel < 0.02 \ihMpc$
inaccessible. As this approach assumes perfect knowledge of the telescope, we
perform a conservative test of how essential this is by simulating and
analysing datasets with deviations about our assumed telescope.  Assuming no
other techniques to mitigate bias are applied, we find we recover unbiased
power spectra when the per-feed beam width to be measured to $0.1 \%$, and
amplifier gains to be known to $1 \%$ within each minute. Finally, as an
example application, we extend our forecasts to a wideband
\SIrange{400}{800}{\mega\hertz} cosmological observation and consider the
implications for probing dark energy, finding a pathfinder-scale medium-sized
cylinder telescope improves the DETF Figure of Merit by around $70 \%$ over
Planck and Stage II experiments alone.

\end{abstract}

\maketitle

\sisetup{detect-none}


\section{Introduction}

Recent years have seen a surge in excitement at the promise of radio
cosmology. By using low frequency observations of the \tcm line we can survey
the distribution of neutral hydrogen throughout large volumes of the
Universe. Radio interferometers provide an efficient, and cost effective
method for doing this.

This transformation of radio interferometers into survey instruments has been
driven by recent technological advances, particularly in the cheap low-noise
amplifiers required for mobile phones, and the constant progress of Moore's law
making large, high bandwidth correlators economical. By correlating a large
number of low cost feeds in a compact area we can produce a telescope ideally
suited for wide-field surveys.

There are three main epochs we can observe: low redshift ($z \lesssim 4$),
where we observe the large scale emission from unresolved galaxies, a technique
termed \emph{intensity mapping} \cite{Chang2008,Wyithe2008}; the \emph{Epoch of
Reionisation} ($z \sim 6$--$10$) where the neutral IGM is eaten away by the
first ionising sources \cite{Furlanetto2006}; and perhaps even the primordial
structure in the \emph{dark ages} ($z \gtrsim 30$), though observations at
these very low frequencies ($\nu < \SI{50}{\mega\hertz}$) will be extremely
challenging \cite{Loeb2004}. These eras are of huge cosmological importance, a
fact reflected in the large number of current and planned experiments targeting
\tcm observations, with
GMRT \cite{GMRT},
HERA \cite{Pober2013},
LOFAR \cite{LOFAR},
MWA \cite{MWA},
MITEoR \cite{MITEoR} and
PAPER \cite{PAPER} targeting the Epoch of Reionisation and 
BAOBAB \cite{BAOBAB},
BAORadio \cite{BAORadio},
BINGO \cite{BINGO},
CHIME \cite{CHIME},
EMBRACE/EMMA \cite{EMBRACE},
Ooty \cite{OotyIM},
Parkes \cite{Parkes} and
Tianlai \cite{TIANLAI} aiming at the low redshift intensity mapping era.

In this paper we will focus on the low redshift, intensity mapping epoch,
though most of the results and techniques we describe apply equally well at higher redshift. Observations
at these low redshifts target the same science as spectroscopic galaxy redshift
surveys such as \cite{WiggleZ,BOSS}: measuring Baryon Acoustic Oscillations and
through them probing the expansion history of the Universe
\cite{Blake2003,Hu2003,Seo2003}. However, they are very complimentary, with
radio observations probing a larger volume at higher redshift, with a
completely different set of systematics.

To make effective use of this new generation of radio interferometers, we must
develop new methods of interpreting and analysing their data. Progress has
accelerated in recent years with many developments
\cite{Myers2003,Tegmark2009,Parsons2009,Liu2010,Liu2011,Parsons2012,Ansari2012,Dillon2012}.

In a previous paper \cite{Shaw2013} we developed a new techique for the
analysis of data from these experiments called the \emph{$m$-mode formalism}.
This method departs from the usual interferometric analysis---making no
flat-sky or small field approximations---at the expense of being limited to
transit telescopes for which it is an exact treatment. It also brings
computational advantages by allowing us to break the data into uncorrelated
$m$-modes, making it feasible to treat the full statistics of the data. This
opens up the possibility of performing optimal map making, foreground
subtraction, and power spectrum estimation, which would be extremely difficult
otherwise.

Perhaps the foremost challenge facing \tcm cosmology is the presence of bright
astrophysical radio sources at frequencies below \SI{1.4}{\giga\hertz} which
are around six orders of magnitude brighter than the \tcm signal.
This emission comes mainly from synchrotron radiation, which is spectrally
smooth, and in principle this allows it to be separated from the \tcm as it is
described by a small number of modes \cite{Liu2012} and these can simply be
removed. The remaining modes, which have significant spectral structure are
assumed to be free of contamination. Unfortunately, this picture is complicated
by the realities of radio observation:
\begin{itemize}
\item Frequency dependent beams lead to mixing of angular structure into
spectral structure which contaminates the foreground clean modes
\cite{Liu2009}. This problem, known as \emph{mode mixing}, means that looking
at \emph{only} the frequency direction of our data is insufficient to separate
these two signals.

\item Synchrotron emission from our galaxy is highly polarised, and though the
totally intensity is spectrally smooth, Faraday rotation by the magnetic
interstellar medium means that the polarised emission is not. Unfortunately,
the complicated polarisation response of real telescopes irreversibly mixes
some fraction of the polarised sky, introducing significant frequency
fluctuations \cite{Moore2013}. As the emission comes from a range of Faraday
depths, we cannot simply de-rotate the emission.

\end{itemize}

Fundamentally there are still the same number of large foreground modes, mode
mixing only makes them harder to identify. In \cite{Shaw2013} we developed a
foreground removal technique based on the \KLfull (KL) transform. This uses the
full covariance statistics of the contaminating foregrounds to find an optimal
separation from the \tcm signal, fully accounting for this mode mixing effect.
However, the technique presented there was limited in two important ways: no
attempt was made to address the problem of polarised foregrounds; and it
assumed that we have full knowledge of the properties of our instrument,
including the full polarised response of the primary beam, and any per-feed
amplitude gains and phase shifts introduced in the receiver system. In this
paper we continue to develop both the $m$-mode formalism and KL transform for
foreground cleaning, with particular emphasis on investigating these two
limitations.

We start by extending the $m$-mode formalism to give a full treatment of
polarisation (\cref{sec:formalism}), and discuss how the unpolarised approach
of \cite{Shaw2013} is a limiting case (\cref{sec:unpolarised})). The example
telescope we use throughout is described in \cref{sec:cylinder}, and its
harmonic space sensitivity is examined in \cref{sec:sensitivity}. Next we take
a careful look at the geometry of the measured $m$-modes (\cref{sec:svd}),
leading us to a technique which both efficiently compresses the data and
effectively removes polarised contamination. We give an overview of the \KLfull
scheme for foreground removal in \cref{sec:kltransform}, and demonstrate its
effectiveness on simulated polarised skies. In \cref{sec:powerspectrum} we
construct an optimal power spectrum estimator in the $m$-mode formalism, which
we use to study the performance of our foreground filter
(\cref{sec:discussion}). In \cref{sec:uncertain} we use this estimator to show
how instrumental uncertainties give rise to power spectrum biases. Finally we
forecast the performance of our example telescope at measuring the expansion
history of the Universe and constraining the nature of dark energy
(\cref{sec:fullbandwidth}).


\section{Polarised Transit Telescope Analysis}
\label{sec:formalism}

In this section we develop a fully polarised version of the $m$-mode formalism,
a new method for analysing transit interferometers that was first introduced in
a previous paper \cite{Shaw2013}. That treatment encapsulates all the essential
ideas but avoids the added complexity of tracking the polarisation, and is a
useful introduction to the full treatment given here. Polarised descriptions of
full-sky interferometry have been given elsewhere (notably
\cite{Kim2007,McEwen2008}), but here we develop the transit telescope limit.

Any transit telescope can be viewed as a collection of feeds, fixed relative
to the ground frame. Each feed, $F_i$ measures a combination of the electric
field $E_a(\vnhat)$ coming from various directions on the sky. In order to
accurately treat the polarisation when the response varies over the sky, we
need to be able to keep track of the contribution from each direction to the
electric field at a point. In order to do this we define $\vEps$ as the
electric field density in a frequency interval $d\nu$ and solid angle $\dhn$
by
\begin{equation}
d\vE = (\mu_0 c)^{1/2} \vEps(\vnhat, \nu)\, \dhn\, d\nu \; .
\end{equation}
With this definition the Poynting flux is conveniently written as
\begin{align}
\vS_p & = \frac{1}{\mu_0 c} \vE \times \vH\notag\\
&= \int \dhn \, \dhn' \, d\nu \, d\nu' \, \vnhat \la \vEps(\vnhat) \cdot \vEps(\vnhat') \ra \; .
\end{align}
Radio emission from the sky is generally incoherent and so we can write the
correlations of $\vEps$ explicitly in terms of the Stokes parameters
\begin{multline}
\label{eq:EEpol}
  \la \eps_a(\vnhat, \nu) \eps_b^* (\vnhat', \nu) \ra = \frac{2 k_B}{\lambda^2} \delta(\vnhat - \vnhat') \delta(\nu - \nu') \\ \times \left[\calP_{ab}^T
    T(\vnhat) + \calP_{ab}^Q Q(\vnhat) + \calP_{ab}^U U(\vnhat) + \calP_{ab}^V V(\vnhat)\right] \; ,
\end{multline}
where the indices are over basis vectors transverse to the line of sight. As
in the unpolarised case we are more interested in the brightness temperature
on the sky, and so we have written \eqref{eq:EEpol} to make that explicit
(thus $Q$, $U$, and $V$ are polarisation brightness temperatures). The polarisation
tensors $\calP^X_{ab}$ are related to the Pauli matrices (in an orthonormal
basis), specifically
\begin{align}
\calP^I_{ab} & = \frac{1}{2}\begin{pmatrix} 1 & 0 \\ 0 & 1\end{pmatrix},
&
\calP^Q_{ab} &= \frac{1}{2}\begin{pmatrix} 1 & 0 \\ 0 & -1\end{pmatrix},
\notag \\
\calP^U_{ab} &= \frac{1}{2}\begin{pmatrix} 0 & 1 \\ 1 & 0\end{pmatrix},
&
\calP^V_{ab} &= \frac{1}{2}\begin{pmatrix} 0 & -i \\ i & 0\end{pmatrix}.
\end{align}
The standard basis to use in spherical geometry are the polar and azimuthal
directions, $\hat{\theta}$ and $\hat{\phi}$, as these allow spin spherical
harmonics to be used straightforwardly to decompose the polarisation field.

Any feed on the telescope measures a weighted combination of the electric
field coming from each direction on the sky. In particular we need to keep
track of the antenna's sensitivity to the orientation of the incoming electric
field. We'll write the measured signal at the $i$-th feed, as $F_i$ which is given by
\begin{equation}
\label{eq:feedresponse}
  F_i(\phi) = \int \dhn \, A^a_i(\vnhat; \phi) \eps_a(\vnhat) e^{2 \pi i
    \vnhat\cdot\vu_i(\phi)}\; ,
\end{equation}
and is directly proportional to the voltage induced in the circuit. Here, and
onwards, we will implicitly sum over the polarisation index $a$. The antenna
reception pattern $A^a_i$ is a vector quantity describing the electric field
response in a given direction. The response $\vA \propto
\vec{l}_\text{eff}$, the effective antenna length (choosing them equal would
make $F_i$ be the antenna voltage). We normalise $\vA$ such that the
normalised antenna power pattern $P_n(\vnhat) = \lv \vA(\vnhat) \rv^2$
ensuring the solid angle of the beam is
\begin{equation}
\Omega_i = \int \dhn \lv \vA(\vnhat) \rv^2 \; .
\end{equation}
In \cref{eq:feedresponse} we have also included an exponential factor which
gives the phase relative to an arbitrary reference point. As both this, and
the antenna orientation change with the Earth's rotation relative to the sky,
we write them as functions of $\phi$, the rotation angle.

The fundamental quantity in radio-interferometry is the cross correlation
between two feeds, the \emph{visibility} $V_{ij} = \la F_i F_j^* \ra$. Using
\eqref{eq:feedresponse,eq:EEpol} we can write down exactly what a
visibility measures, explicitly keeping track of the different sky
polarisations to give
\begin{multline}
\label{eq:vispol}
V_{ij}(\phi) = \int \, \Bigl[ B^T_{ij}(\vnhat; \phi) T(\vnhat) + B^Q_{ij}(\vnhat; \phi) Q(\vnhat)\\
  + B^U_{ij}(\vnhat; \phi) U(\vnhat) + B^V_{ij}(\vnhat; \phi) V(\vnhat)\Bigr] \dhn + n_{ij}(\phi)
\end{multline}
where the beam transfer functions $B^X_{ij}$ encode all the information about
the optics and geometry of the instrument. They are given by
\begin{equation}
\label{eq:beamtransfer_pol}
B^X_{ij}(\vnhat; \phi) = \frac{2}{\Omega_{ij}}A_i^a(\vnhat; \phi) A_j^{b *}(\vnhat; \phi)
\calP^X_{ab} \:e^{2 \pi i \vnhat \cdot
  \vu_{ij}(\phi)} \; ,
\end{equation}
where $\Omega_{ij} = \sqrt{\Omega_i \Omega_j}$. The measured visibilities are
corrupted by noise, which we include as an additional term $n_{ij}$. In this
work we will assume that the noise from different antennas, and frequencies are
uncorrelated (we discuss the statistics in more detail in \cref{app:noise}).

We normalise our visibilities so that they are the correlated antenna
temperature in the noiseless limit (in particular the auto-correlation
\emph{is} the antenna temperature). The factor of two in the definition of the
transfer function ensures that for an unpolarised sky with uniform brightness
$T_b$, the measured autocorrelation $V_{ii} = T_b$. Note that in
\cref{eq:vispol} and onwards the symbol $V$ denotes two different quantities,
the visibility $V_{ij}(\phi)$ and the Stokes V sky field $V(\vnhat)$. The
distinction will be clear from the context.

The above \cref{eq:vispol,eq:beamtransfer_pol} are completely exact. The
general approach to interferometric analysis is to approximate the above to a
2D Fourier transform, which is valid for small fields of view. For wide-field
observations we can attempt to relax this with techniques such as mosaicing
\cite{Holdaway1999} and $w$-projection \cite{Bhatnagar2008}  though this
quickly becomes complicated. In our case we are interested in a specific class
of \emph{transit} interferometers intended for surveys. However, as these
instruments are extremely wide-field, this approach is limiting. Instead we
will try a different route, restricting our domain to transit telescopes, but
otherwise attempting to keep the analysis exact.

To continue, we decompose into spherical harmonics, as they are a natural way
of representing fluctuations on the sky. As polarisation is not a scalar field
we must expand $Q$ and $U$ in spin-2 harmonics $Y_{lm}^\brsc{\pm 2}(\vnhat)$
(the Stokes $V$ field transforms as a scalar). This yields
\begin{align}
T(\vnhat) & = \sum_{lm} a^T_{lm} Y_{lm}(\vnhat) \; ,\\
Q(\vnhat) + i U(\vnhat) & = \sum_{lm} a^\brsc{+2}_{lm}
Y^\brsc{+2}_{lm}(\vnhat) \; ,\\
Q(\vnhat) - i U(\vnhat) & = \sum_{lm} a^\brsc{-2}_{lm}
Y^\brsc{-2}_{lm}(\vnhat) \; , \\
V(\vnhat) & = \sum_{lm} a^V_{lm} Y_{lm}(\vnhat) \; .
\end{align}
The polarised beam transfer matrices also transform as spin fields, and so we
decompose them in the same way, with
\begin{align}
B^T_{ij}(\vnhat; \phi) & = \sum_{lm} B^T_{ij,lm}(\phi) Y_{lm}^*(\vnhat) \; , \\
B^Q_{ij}(\vnhat; \phi) - i B^U_{ij}(\vnhat; \phi) & = \sum_{lm}
B^\brsc{+2}_{ij;lm} (\phi)
Y^{\brsc{+2} *}_{lm}(\vnhat) \; ,\\
B^Q_{ij}(\vnhat; \phi) + i B^U_{ij}(\vnhat; \phi) & = \sum_{lm}
B^\brsc{-2}_{ij;lm} (\phi)
Y^{\brsc{-2} *}_{lm}(\vnhat) \; , \\
B^V_{ij}(\vnhat; \phi) & = \sum_{lm} B^V_{ij,lm}(\phi) Y_{lm}^*(\vnhat) \; .
\end{align}
Note that we have decomposed with the complex conjugates of the spin-harmonics. This allows us to use the orthogonality of the (spin) spherical
harmonics to rewrite the visibility equation \eqref{eq:vispol} as
\begin{multline}
V_{ij}(\phi) = \sum_{lm} \Bigl[B_{ij;lm}^T(\phi) a^{T}_{lm} +
\frac{1}{2} B_{ij;lm}^\brsc{+2}(\phi) a^{\brsc{+2}}_{lm} \\+
\frac{1}{2} B_{ij;lm}^\brsc{-2}(\phi) a^{\brsc{-2}}_{lm} + B_{ij;lm}^V(\phi) a^{V}_{lm}\Bigr] + n_{ij}(\phi)
\; .
\end{multline}
Though this has completely transformed the problem into harmonic space, it
will be more convenient if we change into the conventional $E$ and $B$ mode
decomposition as they are real scalar fields \cite{Zaldarriaga1997}. This can
be done by making the standard substitutions
\begin{align}
a^\brsc{+2}_{lm} & = -\left(a^E_{lm} + i a^B_{lm}\right) \; ,\\
a^\brsc{-2}_{lm} & = -\left(a^E_{lm} - i a^B_{lm}\right)
\end{align}
as well as the corresponding changes for the beam matrices
\begin{align}
  B^\brsc{+2}_{ij;lm} & = -\left(B^E_{ij;lm} - i B^B_{ij;lm}\right) \; ,\\
  B^\brsc{-2}_{ij;lm} & = -\left(B^E_{ij;lm} + i B^B_{ij;lm}\right)
  \; ,
\end{align}
leaving the visibility as
\begin{multline}
\label{eq:vis_prefourier}
V_{ij}(\phi) = \sum_{lm} \Bigl[B_{ij;lm}^T(\phi) a^{T}_{lm} +
B_{ij;lm}^E(\phi) a^{E}_{lm} \\ + B_{ij;lm}^B(\phi)
a^{B}_{lm} + B_{ij;lm}^V(\phi) a^{V}_{lm} \Bigr] + n_{ij}(\phi)
\; .
\end{multline}
In the above the harmonic coefficients are now all the transforms of real
scalar fields (the $B^X_{ij;lm}$ are the complex conjugates of the spherical
harmonic coefficients).

Given the periodicity of the system in $\phi$, Fourier transforming
\cref{eq:vis_prefourier} is an obvious next step
\begin{equation}
V_{ij;m} = \int \frac{d\phi}{2\pi} V_{ij}(\phi) e^{-i m \phi} \; .
\end{equation}
We call these Fourier coefficients, \emph{$m$-modes}, and they will become the
key quantity in our analysis. As the visibility is a \emph{complex} timestream,
the positive and negative $m$'s are independent measurements.

As the $\phi$ dependence simply rotates the functions about the polar axis the
transfer function is trivially $B^{X}_{ij; lm}(\phi) = B^X_{ij;
lm}(\phi\!=\!0) e^{i m \phi}$. The integral over the exponential factors
generates the Kroenecker delta $\delta_{mm'}$ and removes the summation over
$m$ entirely, and we can write the $m$-modes as
\begin{multline}
V_{ij;m} = \sum_{l} \Bigl[B_{ij;lm}^T a^{T}_{lm} +
B_{ij;lm}^E a^{E}_{lm} \\+ B_{ij;lm}^B
a^{B}_{lm} + B_{ij;lm}^V a^{V}_{lm}\Bigr] + n_{ij;m}
\; .
\end{multline}
Though slightly hidden, this a property of the \emph{convolution theorem}. For
a transit telescope the visibility timestream is an azimuthal convolution of
the beam and sky. This means its Fourier conjugate, the $m$-modes, are products
of the individual Fourier modes (with a remaining summation over the $l$
index). This equation fully describes how the measured visibilities are related
to the polarised sky that we are observing.


It is worth thinking about what we are measuring. The visibility we see is a
complex time series, which roughly corresponds to the signal from the sky
modulated by a complex Fourier mode. In our case the time variable is $\phi$,
the Earth's rotation. Taking the Fourier transform of a visibility splits the
time series into right and left moving waves (positive and negative $m$
respectively). A correlated beam pointing south of the north pole only produces
modes moving in one direction (as the beam on the sky is a Fourier mode),
however pointing the same beam beyond the north pole (that is north of it as
defined in the ground frame), produces the other modes as the Fourier mode on
the sky moves in the opposite direction with respect to the Earth rotation. One
important consequence of this is that if we use the freedom to choose the order
of the our feed pairs such that the baseline vectors point towards the east,
positive $m$-modes are produced below the pole, and negative $m$-modes come
from above. If the primary beam does not extend over the pole only positive
$m$'s are produced, though a small amount of negative $m$'s are seen because of
the effect of the primary beam.

In fact, whilst the positive and negative $m$-modes may be independent
measurements they are still observations of the same sky --- for a real field
$a_{lm} = a^*_{l,-m}$ and thus both $V^{ij}_m$ and $V^{ij *}_{-m}$ measure the
same harmonics on the sky. It will be useful to change our notation to make
this fact transparent.

Let us separate out the positive and negative $m$ parts by defining
\begin{align}
B_{ij; lm}^{X, +} & = B_{ij; lm}^X & n^{+}_{ij; m} & = n_{ij; m} \\
B_{ij; lm}^{X, -} & = (-1)^m B^{X *}_{ij; l, -m} & n^{-}_{ij; m} & = n^{*}_{ij;-m}
\end{align}
which is valid for $m \ge 0$. Additionally to prevent double counting the $m =
0$ measurement we need to set $B_{ij; l 0}^{X -} = n_{ij; 0}^{-} = 0$. For
brevity of notation, we will introduce a label $\alpha$ which indexes both the
positive and negative $m$ parts of all included feed pairs $ij$, such that any
particular $\alpha$ specifies exactly the values of $ij,\pm$ (exactly how
$\alpha$ is packed is unimportant). This gives the final form of the $m$-mode
visibility equation that we use as the basis of this work,
\begin{multline}
\label{eq:vis_pol}
V_{\alpha; m} = \sum_{l} \Bigl[B_{\alpha; lm}^T a^{T}_{lm} +
B_{\alpha; lm}^E a^{E}_{lm} \\ + B_{\alpha; lm}^B
a^{B}_{lm} + B_{\alpha; lm}^V a^{V}_{lm} \Bigr] + n_{\alpha; m}
\; .
\end{multline}

As in \cite{Shaw2013} we can write this equation in an explicit matrix form
which will allow us to simplify the notation. The beam transfer matrices above
can be written in an explicit matrix notation
\begin{equation}
\left(\mB^X_m\right)_{(\alpha \nu) (l \nu')} = B^{X, \nu}_{\alpha; m} \delta_{\nu \nu'}
\end{equation}
where the row index labels all baseline ($\alpha$) and frequency combinations
($\nu$), whereas the column index is over all multipole ($l$) and frequencies
($\nu'$). Similarly we can define vectors for the visibilities and harmonic
coefficients
\begin{equation}
\left(\vv_m\right)_{(\alpha \nu)} = V^{\nu}_{\alpha; m} \, \quad
\left(\va^X_m\right)_{(l \nu)} = a_{lm}^{X \nu} \; .
\end{equation}

To keep track of the different polarisation states we define the block matrix
and vector
\begin{equation}
\mB = \left(\begin{array}{c|c|c|c} & & & \\ \mB_T & \mB_E & \mB_B & \mB_V \\ & & &\end{array}\right)
\;, \quad
\va = \left(\begin{array}{c} \va_T \\ \hline \va_E \\ \hline
    \va_B \\ \hline \va_V\end{array}\right)
\end{equation}
such that
\begin{equation}
\label{eq:matnot}
\vv = \mB\, \va + \vn \; .
\end{equation}

This is the essence of the $m$-mode formalism: a simple, linear matrix
relation that exactly describes the whole measurement process for a transit
interferometer. As we will discuss in \cref{sec:sensitivity} both the number
of $m$-modes and the dimensionality of the $\mB$ matrices is bounded by the
physical size of the instrument. This means that we can easily apply all the
standard tools of statistical signal processing without even remembering that
we're dealing with an interferometer. In the following sections we do this
with gusto.

Despite this being an interferometry paper, the $uv$-plane has not been
mentioned at all so far. Though it is prevalent in many interferometric
applications, as both an extremely useful aid for physical understanding and
for computational efficiency (by virtue of the FFT), the $m$-mode formalism
does not make use of it. Eschewing the $uv$-plane is part of its power,
helping it to work trivially for wide-field analysis, and focusing us on only
the measured degrees of freedom. However, it comes at the cost of making it
difficult to have concrete physical interpretations of the process.

\section{Unpolarised Limit}
\label{sec:unpolarised}
In \cite{Shaw2013} we developed an unpolarised formalism because it gives a
simpler problem to analyse, both conceptually and computationally. However,
under certain assumptions it is directly equivalent to the full polarised
case.

For a telescope with dual polarised antennas, let us suppose that we can
engineer our telescope optics such that the field patterns of the two feeds
(labelled $X$ and $Y$) obey two constraints. First, that their normalised
power patterns are equal everywhere
\begin{equation}
\lv \vA_{\scriptscriptstyle X} \rv^2 = \lv \vA_{\scriptscriptstyle Y} \rv^2 = A^2 \; ,
\end{equation}
and second that their polarisation orientations are orthogonal all over the sky
\begin{equation}
\vA_{\scriptscriptstyle X} \cdot \vA_{\scriptscriptstyle Y} = 0 \; .
\end{equation}
Under these constraints there is only one relevant linear combination of the
four $XX$, $YY$, $XY$ and $YX$ visibilities that is sensitive to the total
intensity, the average of the $XX$ and $YY$ visibilities
\begin{align}
V_u & = \frac{1}{2}\lp V_{\scriptscriptstyle X\! X} + V_{\scriptscriptstyle Y\! Y} \rp \\
    & = \frac{1}{\Omega} \int \dhn A^2(\vnhat) e^{2\pi i \vnhat\cdot\vu} T(\vnhat) + \frac{1}{2} \lp n_{\scriptscriptstyle X\! X} + n_{\scriptscriptstyle Y\! Y} \rp \; . \notag 
\end{align}
Because of the properties of the polarisation matrices, this is not sensitive
to the different polarisation modes $Q$, $U$ and $V$, whilst all the
orthogonal combinations are insensitive to the total intensity $T$.

In this limit, the combination $V_u$ is equivalent to the unpolarised
formalism given in \cite{Shaw2013}, if we relabel the noise terms
such that $n = \lp n_{\scriptscriptstyle X\! X} + n_{\scriptscriptstyle Y\! Y}
\rp / 2$. Provided the noise terms are uncorrelated this reduces the power
spectrum down by a factor of two --- that is the unpolarised system
temperature is $T_{\text{sys},u} = T_{\text{sys},p} / \sqrt{2}$.


\section{Cylinder Telescopes}
\label{sec:cylinder}
\begin{figure}
\centering

\includegraphics[width=0.7\linewidth]{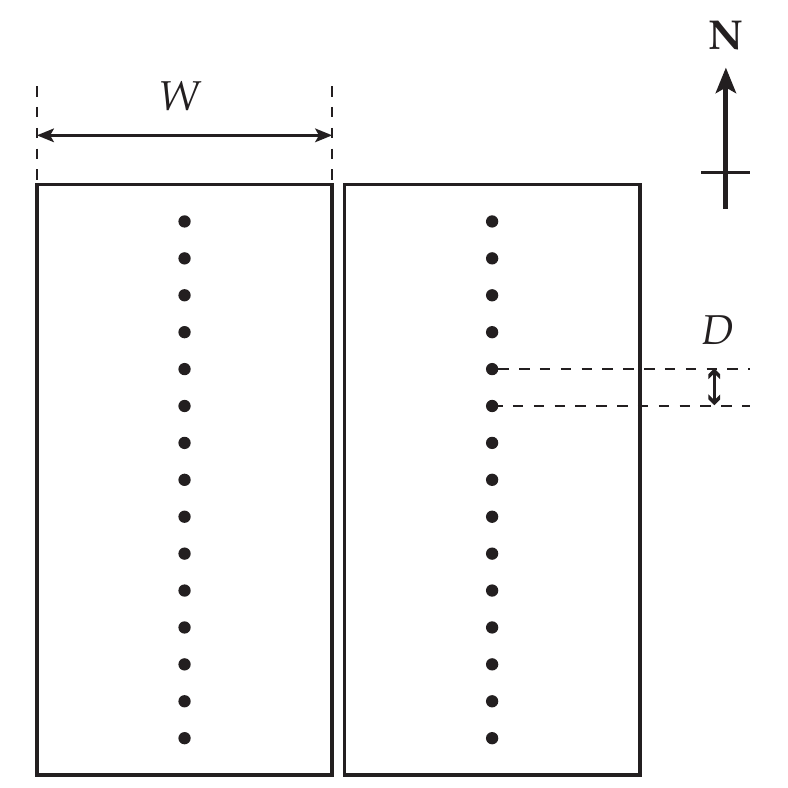}

\caption{A schematic of a cylinder telescope, consisting of two
cylinders aligned North-South on the ground. Each cylinder is of width $W$, and
has $N_\text{feeds}$ regularly spaced a distance $D$ apart. In this paper we
will only consider cylinders which are touching, making the total width of the
array $2 W$. The cylinders are assumed to be long enough that there are no
optical differences between feeds at the edge and in the centre of the array.}

\label{fig:cylinder}
\end{figure}

Cylinder telescopes are interferometric arrays consisting of one or more
parabolic cylindrical reflectors. They have a long history in radio astronomy,
with well known facilities like the Molongo Synthesis Telescope
\cite{Molonglo}, and the Ooty Radio Telescope \cite{Ooty}. Though advances in
amplifier technology meant they steadily lost favour to dish-based
interferometers, interest in them has recently been revived. Reasons are
twofold: the development of cheap, room temperature, low noise amplifiers has
dramatically improved sensitivity; and \tcm intensity mapping has provided an
application for which they are ideally suited.

Intensity mapping requires a large collecting area in a compact region to
achieve high brightness sensitivity, which cylinders can provided cheaply.
Additionally cylinder telescopes are a cost effective way of surveying large
amounts of sky at high speed \cite{Peterson:2006}. And whilst arrays of
dipoles provided a bigger instantaneous field of view, the large number of
elements required at a fixed angular resolution makes the receiver and
correlation hardware increasingly expensive.

Each cylinder has a parabolic cross section such that they focus only in one
direction. In the layout we assume (see \cref{fig:cylinder}), this gives
a long and and thin beam on the sky, extending nearly from horizon to horizon
in the North-South direction but which is only around 2 degrees wide East-West.
Feeds are spaced along the axis of each cylinder --- when correlated these
provide resolution in the N-S direction.  As the telescope operates as a
transit telescope this means that the entire visible sky is observed once per
sidereal day.

In this paper we illustrate the $m$-mode formalism using a medium sized
cylinder telescope, similar to the CHIME Pathfinder. \Cref{tab:cylparams} lists
the parameters of this example instrument.

\begin{table}
\caption{Parameters of the example cylinder telescope.}
\label{tab:cylparams}
\begin{ruledtabular}
\begin{tabular}{ll}
Parameters & Value \\
\hline
Number of cylinders &   2             \\
Cylinder width [m]  &   20            \\
Feeds per cylinder  &   64 (dual-pol) \\
Feed spacing [m]    &   0.3           \\
$T_\text{sys}$ [K]  &   50            \\
Bandwidth [MHz]     &   $400$--$800$  \\
Channel width [MHz] &   2.5           \\
Number of Channels  &   160 (in groups of 40) \\
Telescope Latitude  &   \SI{+45}{\degree}
\end{tabular}
\end{ruledtabular}

\end{table}

\subsection{Beam model}
\label{sec:beammodel}

In the $m$-mode formalism knowledge of the primary beams of our instrument is
crucial. In our model we assume an arrangement such that at each location
there are two perpendicular dipoles: the $X$ feed where the dipole is aligned
across the cylinder (pointing East), and $Y$ feed where the dipole is aligned
along the axis (pointing North). In both cases the feeds hang below a
conducting ground plane which stops the beam spilling above the cylinder
(which is assumed to have an f-ratio of $1/4$).

Solving for the beam on the sky for a feed placed in a parabolic cylinder is a
complex problem (for one approach see \cite{Craeye2005,Craeye2008}). Crudely
the cylinder acts in two ways: in the parabolic direction it focuses the
antenna beam to a diffraction limited beam on the sky; in the orthogonal
direction it acts like a mirror, inverting the antenna beam. Rather than try
to accurately solve for the beam, we try to capture these two effects. We will
break the model down into the product of two 1D functions: a function for the
E-W direction, calculated by illuminating the cylinder with the dipole beam,
and solving for the diffraction in the Fraunhofer limit; and a N-S function
which is just the reflected feed amplitude in the N-S direction. We will also
model the polarisation direction as being the same as that of an unfocused
dipole (in spherical co-ordinates, for a dipole along the polar axis, the
polarisation direction is $\hat{\vec{\theta}}$).

First we model the beam amplitude for the unfocused dipole in the E-plane and
H-plane as taking the form
\begin{equation}
A_D(\theta; \, \theta_W) = \exp{\lp - \frac{\ln{2}}{2} \frac{\tan^2\theta}{\tan^2\theta_W}\rp} \; ,
\end{equation}
where $\theta_W$ is the full width at \emph{half-power} of the beam. For a
horizontal dipole mounted a distance $\lambda / 4$ over a conducting ground
plane (see \cite[section 4.7]{AntennaTheory}), we can exactly calculate the
widths in the H-plane ($\theta_H = 2\pi / 3$) and E-plane ($\theta_E \approx
0.675 \theta_H$). We use these value for our fiducial beam model, though we
will vary them later in this paper.

In the E-W direction we are solving the Frauhofer diffraction problem of a
cylinder feed illuminating an aperture of a finite width. This has the
solution
\begin{align}
A_F(\theta; \theta_W, W) & \propto \int_{-\frac{W}{2}}^{\frac{W}{2}}\! A_D(2\tan^{-1}({\scriptstyle\frac{2 x}{W}}); \theta_W) e^{-i k x \sin\theta} dx \notag \\
& \propto \int_{-1}^{1} e^{-\frac{\ln{2}}{\tan^2\theta_W}\frac{u^2}{1 - u^2} -i \frac{\pi W}{\lambda} u \sin{\theta}} du
\end{align}
where we have used the fact that for a cylinder with an f-ratio of $1/4$ a ray
striking a distance $x$ from the cylinder centre reflects by an angle $\theta
= 2 \tan^{-1}{\lp 2 x / W\rp} = 2\tan^{-1}{u}$ where W is the cylinder width.

Putting these components together, our overall beam model can be written as
the product of three functions. For the $X$ feed
\begin{multline}
A_a^X(\vnhat) = A_F(\sin^{-1}(\vnhat\cdot\vxhat); \,\theta_E, W) \\ \times A_D(\sin^{-1}(\vnhat\cdot\vyhat); \,\theta_H)  p_a(\vnhat;\, \vxhat) 
\end{multline}
and for the $Y$ feed
\begin{multline}
A_a^Y(\vnhat) = A_F(\sin^{-1}(\vnhat\cdot\vxhat); \,\theta_H, W) \\ \times A_D(\sin^{-1}(\vnhat\cdot\vyhat); \,\theta_E)  p_a(\vnhat;\, \vyhat) 
\end{multline}
where the vectors $\vxhat$ is a unit vector transverse to the cylinder,
pointing East, and $\vyhat$ is along the cylinder, pointing North. The function
$p_a$ gives the unit vector polarisation direction on the sky for a dipole in
direction $\vdhat$
\begin{equation}
\hat{p}_a(\vnhat; \vdhat) = \frac{1}{\bigl( 1 - (\vnhat\cdot\vdhat)^2 \bigr)^{1/2}} \bigl[ \vdhat - (\vnhat\cdot\vdhat) \vnhat \bigr]_a \; .
\end{equation}

\begin{figure}

\includegraphics[width=\linewidth]{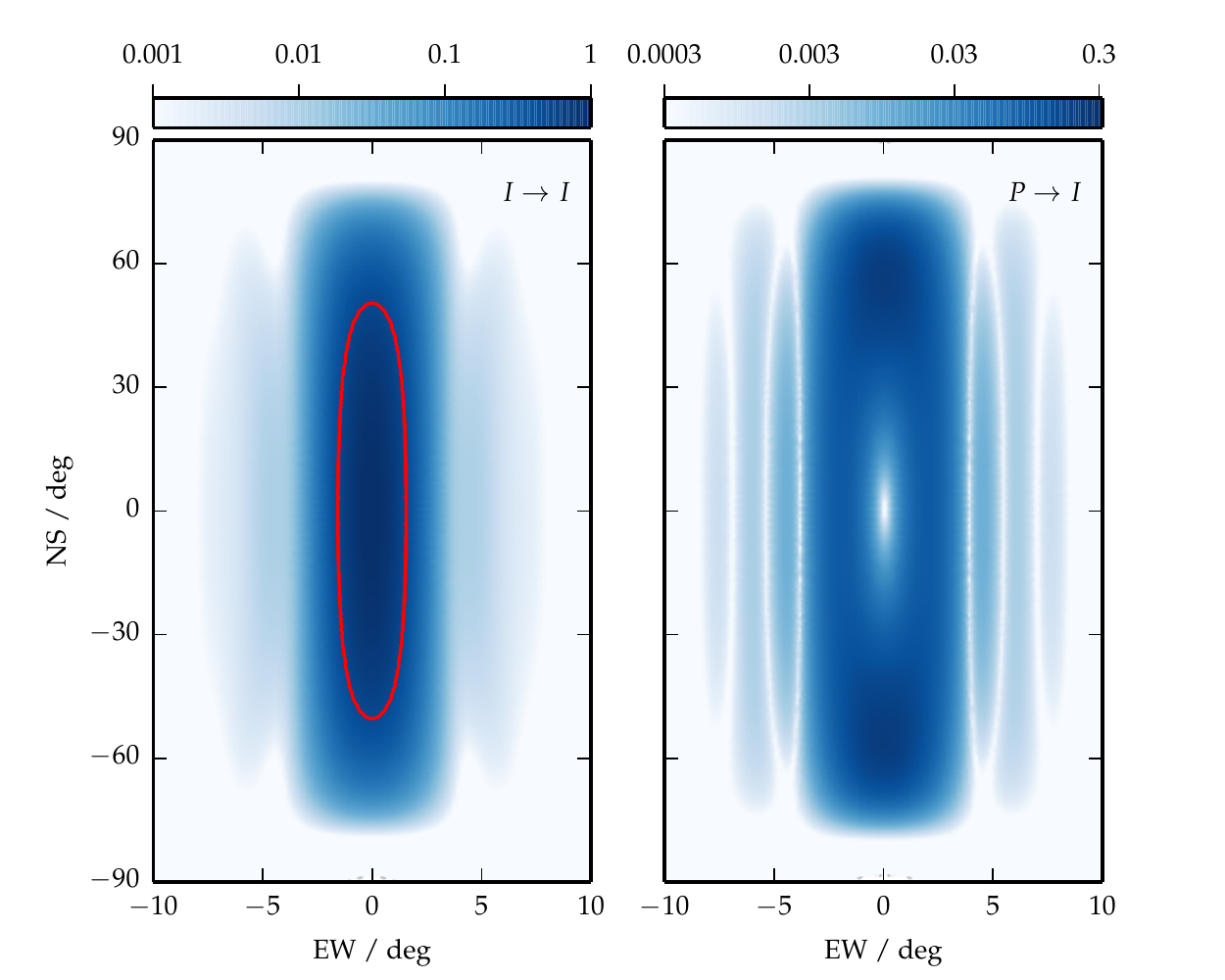}

\caption{The primary beam of the cylinder telescope forms a long strip on the
sky from North to South. This figure illustrates the transfer into an
instrumental Stokes I ($XX + YY$ polarisations), from the total intensity on
the sky (left panel), and from the polarised sky only (right panel). The red
contour in the top panel, marks the half power point of the beam.}
\label{fig:beam}
\end{figure}

In \cref{fig:beam} we illustrate the on-sky beam for this our example
telescope. We plot the response of an `instrumental Stokes I', constructed
from the combination of $XX + YY$ polarisation, to Stokes I and polarisated
emission on the sky. The reponse to Stokes I on the sky is given by
\begin{equation}
\label{eq:rII}
R_{I \rightarrow I} = \lp A^a_X A^b_X + A^a_Y A^b_Y \rp \mathcal{P}_{ab}^I \; .
\end{equation}
As a measure of the response to polarised radiation we use
\begin{equation}
\label{eq:rPI}
R_{P \rightarrow I}^2 = \sum_{P \in \{Q,U,V\}} \bigl[ \lp A^a_X A^b_X + A^a_Y A^b_Y \rp \mathcal{P}_{ab}^P \bigr]^2 \; .
\end{equation}
For a beam with no polarisation leakage, this response is zero. Though our
example has no leakage on-axis, \cref{fig:beam} clearly shows that there is
significant pickup of polarisation away from the beam centre.

\section{Sensitivity in Harmonic Space}
\label{sec:sensitivity}

The geometry of an interferometer on the ground determines its angular
sensitivity on the sky, with the total size of the optical system determining
the smallest scales that can be measured. This limits the number of harmonic
modes on the sky that we are able to measure, reducing \cref{eq:vis_pol} to
finite sums.

The set of spherical harmonics that a given baseline is sensitive to can be
found by expanding a plane wave on the sky
\begin{equation}
\label{eq:planewave_exp}
e^{2\pi i \vnhat \cdot \vu} = \sum_{lm} \ls 4\pi i^l j_l\lp 2\pi \lv \vu \rv \rp Y^*_{lm}(\vuhat) \rs Y_{lm}(\vnhat)
\end{equation}
where the part in square brackets is the coefficient in a spherical harmonic
expansion of the plane wave. The amplitude of a spherical harmonic function can be
conveniently written in terms of integrals of Bessel functions \cite[section
5.4]{Varshalovich}. For large $l \gg 1$ we find
\begin{align}
\lv Y_{lm}(\theta, \phi) \rv^2 & = \frac{2 l + 1}{4\pi} \int_0^\infty \!\! \ls J_m\lp\frac{t \sin{\theta}}{2}\rp \rs^2 \!\! J_{2 l + 1}(t) \, dt \notag \\
& \approx \frac{l}{2\pi} J_m\lp l \sin{\theta} \rp^2
\end{align}
where we have used the approximation that $\lim_{n \rightarrow \infty} J_n(x)
= \delta(x - n)$. Combining this with \cref{eq:planewave_exp} shows that the
magnitude of the spherical harmonic coefficients of a plane wave are
\begin{equation}
\label{eq:plane_exp}
\lv a_{lm} \rv^2 = 8\pi\, l\, j_l(2\pi \lv u \rv)^2 J_m\lp l \sin{\theta} \rp^2 \; .
\end{equation}
In particular this shows that the coefficients are effectively bounded in a
triangle by $l < 2\pi \lv \vu \rv$ and $ \lv m \rv < l \sin{\theta}$ because of
the exponential decay of the Bessel functions for large order.

The highest frequency fourier mode measured on a sky by an individual baseline
comes from the maximum distance between illuminated areas on the correlated
antennas (this is the largest distance from the origin in the $uv$-plane).
Following through from \cref{eq:plane_exp}, we expect the range of
measureable modes to be $l < 2\pi d_\text{max} / \lambda$ and $m < 2\pi
d_\text{E-W} / \lambda$, where $d_\text{max}$ is the largest distance assoicated with
baselines and $d_\text{E-W}$ the largest in the E-W direction.

Let us consider our cylinder (see \cref{fig:cylinder}). A feed on the
cylinder effectively illuminates the whole width of the cylinder, but a very
short distance along its axis. This makes the largest E-W distance of all feed
pairs $N_\text{cyl} W$, and the largest N-S distance $N_\text{feeds} D$. In
terms of spherical harmonics coefficients on the sky, we are limited to
\begin{align}
l & < \frac{2\pi}{\lambda} \sqrt{(N_\text{cyl} W)^2 + (N_\text{feeds} D)^2} \; , \\
m & < \frac{2\pi}{\lambda} N_\text{cyl} W \; .
\end{align}

Though this result is correct for a cylinder telescope, for an interferometer
with a compact field of view, pointing away from the celestial equator, it
needs modifying. As before the resolution in the E-W direction is determined by
the maximum distance $d_\text{E-W}$, however, if the primary beam does not
cross the equator this resolution corresponds to a larger fraction of the
circle of constant declination at that point. As the $m$-mode corresponds to
the Fourier mode in the azimuthal direction, this means that the limit on $m$
is in fact $m < 2\pi \cos{\delta} \: d_\text{E-W} / \lambda$, where $\delta$ is
the declination of the point in the primary beam closest to the celestial
equator.

To look at the sensitivity of the telescope in more detail we can calculate
the Fisher matrix of the $a_{lm}$ coefficients (we discuss the interpretation of Fisher matrices in detail in \cref{sec:powerspectrum}). For Gaussian noise the likelihood function for the $a_{lm}$'s is
\begin{equation}
\mathcal{L}(\va; \vv) \propto \exp{\lp -\frac{1}{2} \lp \vv - \mB \va \rp^\hconj \mN^{-1} \lp \vv - \mB \va \rp \rp}\; .
\end{equation}
From this we can calculate the Fisher matrix for a particular $m$
\begin{align}
\mathcal{F}_{l l'} & = -\la \frac{\partial^2}{\partial a^T_{l} \partial a^T_{l'}} \ln{\mathcal{L}} \ra \notag \\
& = \ls \mB_T^\hconj \mN^{-1} \mB_T  \rs_{l l'} \; .
\end{align}
We expect that in general this matrix will be singular and hence we cannot
find the covariance matrix of the $a_{lm}$ coefficients by finding
$\mat{\mathcal{F}}^{-1}$. One obvious source of this is that the
interferometer does not see the whole sky --- anything declination less than
$\delta = \SI{-45}{\degree}$ is below the horizon --- and this manifests
itself as correlated combinations of $a_{lm}$'s that we cannot separate. Additionally the angular resolution falls off towards the horizon meaning that we do not have uniform sensitivity across the sky.

This consequence of this is obvious from simply counting the degrees of freedom
involved. For the example telescope, there are 762 unique baselines each of
which gives a noisy complex measurement of the sky. However, we are sensitive
up to $l_\text{max} \sim 400$ for each polarisation, giving $4(l_\text{max} -
m)$ complex degrees of freedom on the sky, so there must be some combinations
about which we have no information.

\begin{figure}

\includegraphics[width=\linewidth]{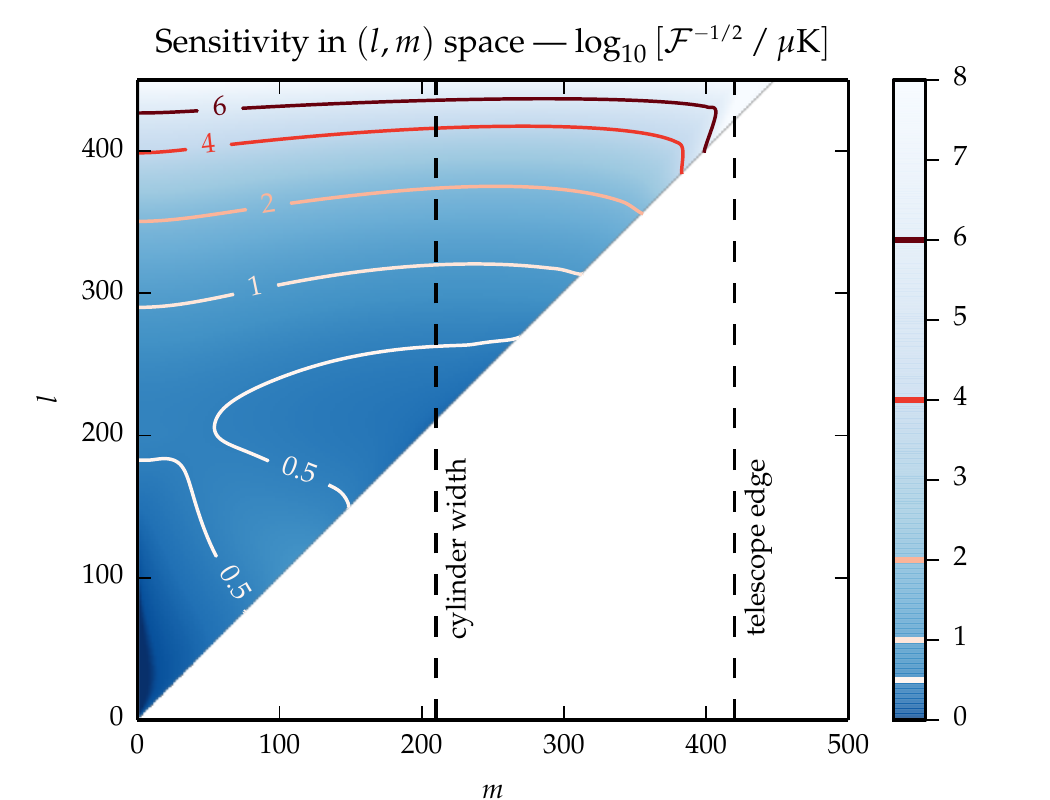}

\caption{Sensitivity of the array to temperature, derived from the inverse of
the diagonal elements of the Fisher matrix $(\mathcal{F}_{(lm)(lm)})^{-1}$.
The plot above shows the $\log_{10}$ of the sensitivity in units of
\si{\micro\kelvin}. The sensitivity to the three remaining Stokes parameters
are largely identical. The dashed black lines mark the $m$ corresponding to
the separation between the cylinders, and the total width of the cylinders. As
we would expect the sensitivity peaks in $m$ at the zero separation, and the
single cylinder separation. It then falls off rapidly at the edge of the
telescope.}

\label{fig:fisher_sensitivity}
\end{figure}

As in general we cannot determine the covariance matrix of the $a_{lm}$, we
will use the Fisher matrix itself to describe the sensitivity. In
\cref{fig:fisher_sensitivity} we show the diagonal elements of the Fisher
matrix at each $m$ for the example telescope, this gives an illustration of
the amount of information we have about any spherical harmonic mode.

\section{SVD projection}
\label{sec:svd}

\begin{figure}

\begin{center}
\includegraphics[width=0.9\linewidth]{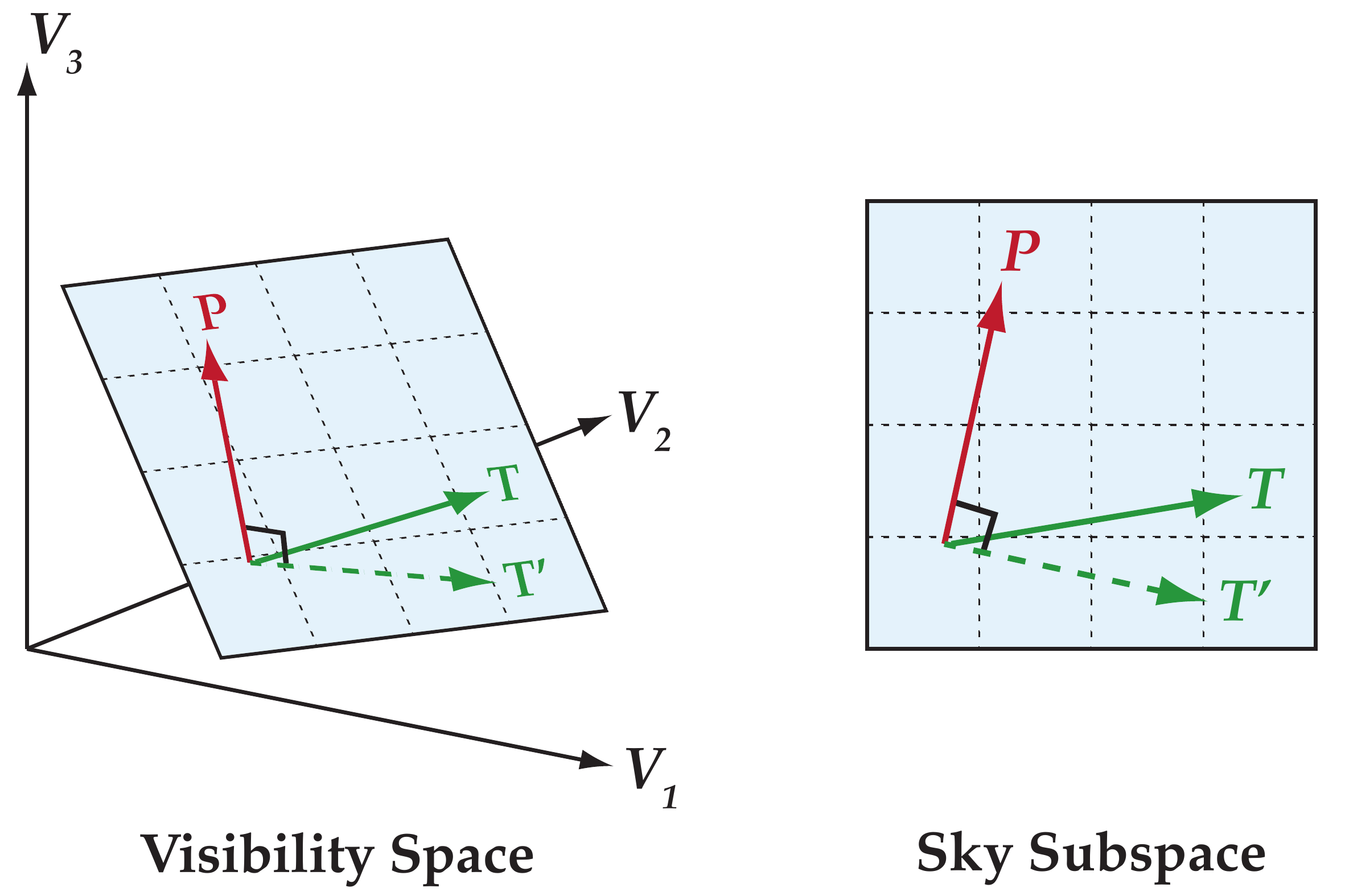}
\end{center}

\caption{The information about the sky does not spread throughout the space of
visibilities but is contained in a subspace, a linear combination of the
measured signals which does not span the whole visibility space. Directions
orthogonal to this subspace are excited only by the instrumental noise, and
contain no information about the sky. The left panel illustrates the geometry
of the full visibility space, showing the sky subspace as a plane. In the right
panel we show only the sky plane. Within this sky subspace, there are yet lower
dimensional subspaces that the total intensity (labelled $T$) and polarised
($P$) signals get mapped to. However, they need not be orthogonal, an effect we
must take into account. One way of treating this is to project onto the space
orthogonal to polarisation (labelled $T'$), this eliminates polarised
contamination at the expense of some sensitivity to total intensity. This is
discussed in detail in the text.}

\label{fig:svdgeometry}
\end{figure}

For \tcm cosmology we are only interested in deriving real properties of the
\emph{unpolarised} sky. As we shall see this is usually of much lower
dimension than the space of measurements made by an interferometer, leaving a
large number of redundant degrees of freedom which are just filled by the
instrumental noise. Eliminating these would allow us to significantly compress
the data space, without losing useful information. In \cref{fig:svdgeometry}
we illustrate the geometry of the measured visibilities. The matrix $\mB$
wholly describes the mapping between the sky and the measured visibilities,
and understanding its structure is the key to isolating the important degrees
of freedom.

To start with let us concentrate on how to reduce to only the degrees of
freedom on the sky (ignoring their polarisations for now). The matrix $\mB$
tells us how a subspace of the spherical harmonics $\va$ map into a subspace
in visibility space $\vv$. This visibility subspace (shown by the plane in the
\cref{fig:svdgeometry}), is termed the \emph{image} of $\mB$. The subspace of
visibilities orthogonal to the image, is called the \emph{cokernel}. The
cokernel has no mapping to the sky, and so measuring this subspace yields no
useful information. By projecting our data onto the image, we remove the
cokernel and compress our data by retaining only the relevant degrees of
freedom. In \cref{fig:svdgeometry} this corresponds to projecting onto the
plane, eliminating the perpendicular dimensions.

The number of retained degrees of freedom is given by the dimensionality of
the image --- that is, the \emph{rank} of $\mB$ --- and cannot exceed the
number of measured modes on the sky. For a single frequency and $m$, the rank
is guaranteed to be less than the total number of spherical harmonics required
to describe the polarised sky, that is $4 (l_\text{max} - m)$. However, in the
case of incomplete sky coverage, we cannot measure all spherical harmonic
modes independently, and this coupling means that the rank is likely to be
reduced to around $4 f_\text{sky} (l_\text{max} - m)$, where $f_\text{sky}$ is
the fraction of sky observed.

These numbers depend only on the physical size of the telescope, and not
details of the feed distribution. For compact interferometers with little
redundancy, the number of feed pairs rapidly exceeds the rank of the matrix,
and so projecting onto the image gives a large computational saving.



To find the image of $\mB$ we can use the Singular Value Decomposition (SVD).
However, first we will pre-whiten the visibilities with respect to the
instrumental noise. This transforms it to be uncorrelated with unit variance
and can be done by multiplying them with $\mNh$, a matrix such that $\mNh
(\mNh)^\hconj = \mN^{-1}$. As $\mN$ is positive definite this factorisation
always exists and can be found by Cholesky factorisation or eigendecomposition.
This leaves \cref{eq:matnot} as
\begin{equation}
\mNh \vv = \mNh \mB \va + \mNh \vn \; .
\end{equation}
%
%
We then take the SVD of the whitened beam transfer matrix
\begin{equation}
\mNh \mB = \mU \msigma \mV^\hconj \; .
\end{equation}
The matrix $\mU$ defines the image and cokernel, given by columns of $\mU$
corresponding to non-zero and zero singular values respectively. In practice
many singular values are numerically small but not precisely zero, giving
modes which are either non-zero because of numerical precision, or simply
carry a very small but non-zero amount of information about the sky. In this
case we separate the image and cokernel using a numerical threshold. We
partition the columns of the matrix $\mU$ into two matrices $\mU_I$ and
$\mU_N$ which give the image and cokernel respectively. To compress our data
we simply filter with the matrix $\mU_I$ to give $\vv' = \mU_I^\hconj \mNh
\vv$.

While this filtering can yield a large compression, we should note that it
preserves all the information about the sky. However, the cosmological signal
we are interested in is purely unpolarised and requires only $\sim f_\text{sky}
(l_\text{max} - m)$ modes per frequency and $m$ to describe it. This suggests
that we should be able to improve our compression by around another factor of
four.

As a first attempt we might consider projecting onto the image of $\mB_T$, the
total intensity transfer matrix, rather than the full $\mB$. In
\cref{fig:svdgeometry} this would correspond to projecting straight onto the
$T$ vector, rather than just the plane.

Unfortunately as illustrated in \cref{fig:svdgeometry}, the image of the total
intensity need not be orthogonal to the subspace containing the polarised
image. This is a manifestation of polarisation leakage. In this case by doing
this we lose the ability to differentiate between polarised and unpolarised
signals from the sky, resulting in catastrophic leakage of polarised
foregrounds.

A resolution to this problem is to project not onto the image of $\mB_T$ but
to perform another projection, this time onto the polarisation cokernel. In
\cref{fig:svdgeometry} this is equivalent to projecting onto the vector $T'$.
By doing this we ensure that there is no leakage of the polarised sky into our
compressed data, at the expense of throwing away information about the total
intensity signal that lies in the overlap between the two spaces.

To project out the polarised signal, we first construct the polarisation
transfer matrix
\begin{equation}
\mB_\text{pol} = \left(\begin{array}{c|c|c} & & \\ \mB_E & \mB_B & \mB_V \\ & & \end{array}\right) \; ,
\end{equation}
then we use this to isolate the polarisation cokernel in the sky compressed
basis by performing another SVD
\begin{equation}
\mU_I^\hconj \mNh \mB_\text{pol} = \mU_\text{pol} \msigma_\text{pol} \mV_\text{pol}^\hconj \; .
\end{equation}
As before we separate into the image and cokernel of this matrix, by dividing
up $\mU_\text{pol}$ into $\mU_{\text{pol},I}$ and $\mU_{\text{pol},N}$
respectively. As before the separation onto the two spaces is not exact, but
done through a numerical threshold. By projecting our dataspace onto the
cokernel we achieve this final compression.

\begin{figure}
\includegraphics[width=\linewidth]{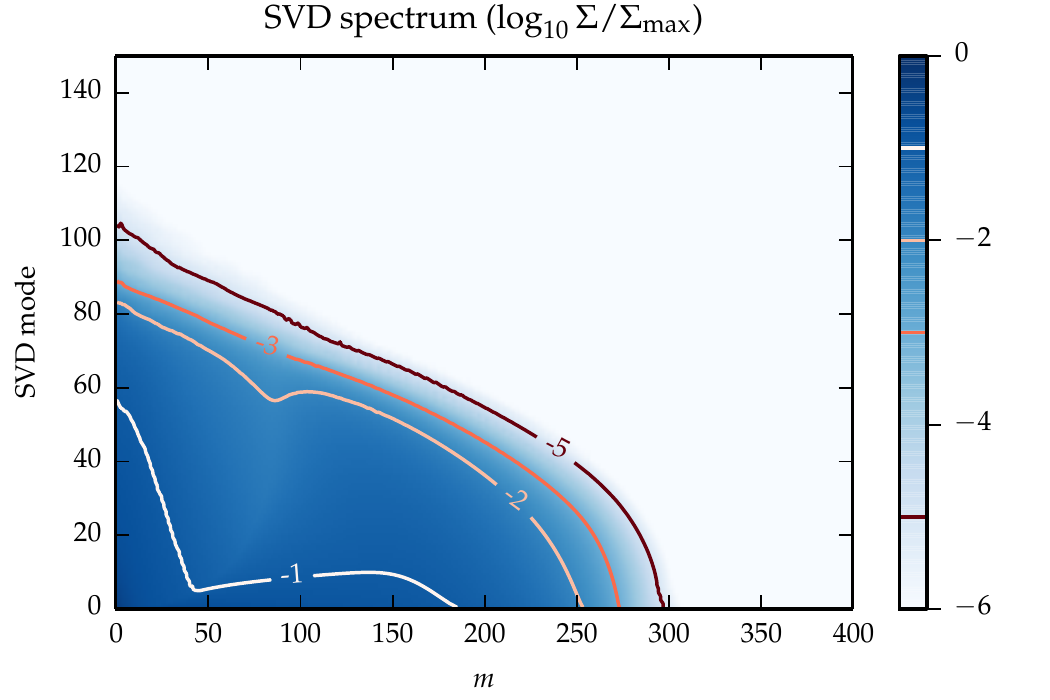}
\caption{The singular values of $\mBb$ for the \SI{400}{\mega\hertz} channel
after removal of the polarised modes. Large singular values represent modes on
the sky that are well measured. We see that at each $m$ there are less than
$100$ measured degrees of freedom from the sky, with the spectrum dropping off
very steeply beyond this. This is a significant saving, before compression
there are twice as many modes as there are unique baselines, including positive
and negative $m$. In our example there are $762$ unique baselines (without
auto-correlations), so there would be $\sim 1500$ modes.}
\label{fig:svdspectrum}
\end{figure}

Overall we have applied three transformations to our data:
\begin{itemize}
\item Whiten the instrumental noise by applying $\mNh$
\item Project onto the sky subspace by using $\mU_I^\hconj$
\item Project out the polarised sky using $\mU_{\text{pol}, N}^\hconj$
\end{itemize}
Combined these define a new basis in which to consider our data. One which
strives to preserve as much of the relevant information as possible, whilst
vastly reducing the number of degrees of freedom we must consider. We define
our filtered visibility data as
\begin{equation}
\vvb = \mU_{\text{pol}, N}^\hconj \mU_I^\hconj \mNh \vv \; .
\end{equation}
We can write a modified version of the measurement equation (\ref{eq:matnot})
which relates this to the sky signal
\begin{equation}
\label{eq:matnot_svd}
\vvb = \mBb\, \va + \vnb
\end{equation}
where we have defined
\begin{align}
\mBb & = \mU_{\text{pol}, N}^\hconj \mU_I^\hconj \mNh \mB \; , \\
\vnb & = \mU_{\text{pol}, N}^\hconj \mU_I^\hconj \mNh \vn \; .
\end{align}
As the columns of $\mU_I$ and $\mU_{\text{pol}, N}$ are orthonormal, the
instrumental noise still has the identity covariance $\langle \vnb \vnb^\hconj
\rangle = \mNb = \mI$. In \cref{fig:svdspectrum} we show the singular values of
the new mapping matrix $\mBb$, clearly illustrating that we are only sensitive
to a small number of modes on the sky.

In order to visualise our data we will want to make maps from our filtered
dataset. For Gaussian distributed instrumental noise it is straightforward to
make maximum likelihood maps of the sky as discussed in \cite{Shaw2013}. As we
have whitened the instrumental noise, our data has a likelihood function
\begin{equation}
\mathcal{L}(\va; \vvt) \propto \exp{\lp -\frac{1}{2} \lv \vvb - \mBb \va \rv^2 \rp}
\end{equation}
and thus we can solve for the maximum-likelihood solution using the
Moore-Penrose pseudo-inverse, giving our best estimate of the spherical
harmonics simply as
\begin{equation}
\label{eq:vis_svd}
\hat{\va} = \mBb^+ \vvb \; .
\end{equation}
As in \cite{Shaw2013}, to make a full map of the sky, we simply use this
estimator on a per-$m$ and per frequency basis and collate the estimates. We
can then perform an inverse Spherical Harmonic Transform to produce sky maps
at each frequency. As we have projected onto the polarisation cokernel, the
data does not contain any information about the polarised sky. Combined with
the minimum power property of the Moore-Penrose pseudo inverse this means that
the polarised spherical harmonics will be zero.

\section{Foreground Removal with the \KLfull Transform}
\label{sec:kltransform}
The foremost challenge for any \tcm intensity mapping experiment is separating
the cosmological signal from astrophysical contaminants which are around
$10^4$--$10^6$ times larger. The primary sources are the diffuse synchrotron
emission from our own galaxy and emission from extra-galactic point sources
\cite{SantosCoorayKnox}. All significant foregrounds are expected to be
spectrally smooth \cite{Liu2012}, however, the \tcm signal decorrelates
quickly as each frequency corresponds to a different spatial slice. This gives
an opportunity to separate the two.

Conceptually foreground removal is simple---we just remove the smooth
frequency component from our observations. Unfortunately the reality is far
from straightforward. The large dynamic range between the amplitude of the
foregrounds and the \tcm signal makes several effects extremely problematic.

\begin{description}

\item[Mode mixing] \hfill \\ In a real experiment the shape of the beam
on the sky will vary with the observed frequency, driven by the optical effects
of using a fixed physical aperture or feed spacing. Even if the angular
fluctuations on the sky were frequency independent as we scan through in
frequency the beam structure changes, and this introduces variations of our
measurements with frequency.

\item[Model uncertainties] \hfill \\ Astrophysical foregrounds are poorly
constrained at the small angular and frequency scales that will be probed by
upcoming \tcm intensity mapping experiments. Whilst there exist theoretical and
phenomenological models of this regime, a successful foreground removal method
should be robust to uncertainties in the foreground statistics. Though most
effort has focused on the uncertainties in the two-point correlations, we must
also ensure that higher order moments do not impair our analysis.

\end{description}

Given these complications, we would prefer a foreground removal method to be
conservative, throwing away potentially useful information in order to be
robust to them. It is better to be cautiously correct than precisely wrong.

Accepting that we may prefer to lose information about the \tcm signal in
order to be unbiased by residual foregrounds, we would still like to perform
the best job we can, requiring that we are
\begin{description}

\item[Statistically Optimal] \hfill \\ Whatever space the foregrounds are
removed in we must be able to keep track of the statistics of both the
instrumental noise, and the foreground residuals in order to be able to
optimally perform subsequent stages, notably power spectrum estimation.

\end{description}
This latter point is especially pertinent for any technique that operates
directly in map space. It is not only difficult to express the pixel-pixel
correlations caused by the measurement process (especially with noise), but
similarly difficult to project these back after any foreground cleaning has
been performed.

In a previous paper \cite{Shaw2013} we developed a foreground removal
technique that addresses these three issues. It does this by explicitly taking
into account the statistics of both the signal and foregrounds in the basis
that they are measured. In this Section we give an overview of this method in
the context of the polarised analysis presented here.

\subsection{Stationary Statistics}
Understanding the statistics of our measured data is essential ingredient in
all but the most basic analysis if we make best use of the data. For intensity
mapping experiments, our data has three components: the \tcm signal which we
are trying to extract, the foregrounds, and instrumental noise. The statistics
of instrumental noise live in the visibility space, the basis of our
measurements. However the other components are naturally represented on the
sky, and must be projected into this space using \cref{eq:vis_svd}.

In this work we treat the sky as a statistically isotropic field with a
two-point function
\begin{equation}
\la a_{l m \nu'} a_{l' m' \nu'}^{*} \ra = C_l(\nu, \nu') \delta_{l l'}
\delta_{m m'} \; ,
\end{equation}
which we write in matrix form as $\mC_\text{sky}$ defined as
\begin{equation}
\ls \mC_\text{sky} \rs_{(l \nu)(l' \nu')} = C_l(\nu, \nu') \delta_{l l'} \; .
\end{equation}
This quantity can be projected into the SVD basis for a given $m$ using the
transfer matrix $\mBb$, which means the final two-point function can be
written as
\begin{equation}
\mCb = \mBb \mC_\text{sky} \mBb^\hconj + \mNb \; .
\end{equation}

As the measurement process itself does not mix $m$-modes, provided the
statistics of the sky do not couple them (which is the case for a statistically
isotropic sky), then the covariance of the data is block diagonal in $m$. This
brings huge computational savings, and makes a full analysis tractable
\cite{Shaw2013}. Clearly the observed sky is not statistically isotropic, with
our own galaxy varying wildly across the sky. However, as discussed in
\cite{Shaw2013}, this does not seem to diminish the effectiveness of the
analysis.

These savings come because we can then operate on each block independently. For
instance to diagonalise a covariance (an $O(N^3)$ operation) we can save around
a factor of $m_\text{max}^2$ in computation by diagonalising each block
separately, and as we only need store the diagonal blocks storage is reduced by
a factor of $m_\text{max}$.

\subsection{Foreground Removal}

Any foreground removal method aims to find a subset of the data within which
there is significantly more \tcm signal than astrophysical foregrounds. Most
techniques are linear, and they can be thought of as constructing a new linear
basis for the data which localises the two components into distinct regions.
Unfortunately, in the presence of mode-mixing, it is not obvious how to select
a basis which separates the two components --- what we introduce here is a
method which can automatically generate it.

The signal covariances of the signal and foregrounds describe how their
respective power is distributed and correlated within the measured data. It is
these correlations that make the foreground fluctuations superficially seem
much larger than those of the signal. In fact we expect them to be driven by a
very small number of very highly correlated modes, and we would like to change
to a basis where this is apparent. This can be achieved by use of the \KLfull
transform (often called the Signal-Noise eigen-decomposition), which has a long
history in cosmology \cite{Bond1994,Tegmark1997b,Bunn2003}. This transform
simultaneously diagonalises both the signal and foreground covariance matrices,
generating an uncorrelated set of modes. This makes comparing the amount of
signal and foreground power in each mode trivial.

Performing this transform requires covariance matrices for the signal and
foregrounds. The signal matrix, $\mS$ contains only the 21-cm signal the we
want to extract
\begin{equation}
\mSb = \mBb \mC_{21} \mBb^\hconj
\end{equation}
whereas the noise covariance contains the astrophysical foregrounds
\begin{equation}
\mFb = \mBb \mC_f \mBb^\hconj \; .
\end{equation}
This requires models for the statistics of both the signal and the
foregrounds. The signal is modelled as a simple Gaussian random field for the
\tcm emission from unresolved galaxies, whereas the foreground model includes
both the synchrotron emission from our galaxy, and the contribution from a
background of extragalactic point sources. The details of both are discussed
in
\cref{app:models}.

Using these two matrices we can construct the \KLfull eigenbasis (see
\cref{app:snmodes} for details on the process). This gives a set of
statistically uncorrelated eigenmodes, and corresponding eigenvalues. Writing
the eigenvectors in a matrix row-wise gives the transformation matrix to
diagonalise the covariances. By convention the signal covariance transforms to
\begin{equation}
\mSb \rightarrow \mSb' = \mP \mSb \mP^\hconj = \mLambda \;,
\end{equation}
where $\mLambda$ is the diagonal matrix of eigenvalues, and the foreground
covariance becomes
\begin{equation}
\mFb \rightarrow \mFb' = \mP \mF \mP^\hconj = \mI \; .
\end{equation}
Hence, in the new basis the eigenvalues $\lambda$ give the ratio of signal to
foreground power. In practice the S/F spectrum is steep, with a quick
transition from foreground dominated to signal dominated modes \cite{Shaw2013}.

\begin{figure}

\includegraphics[width=\linewidth]{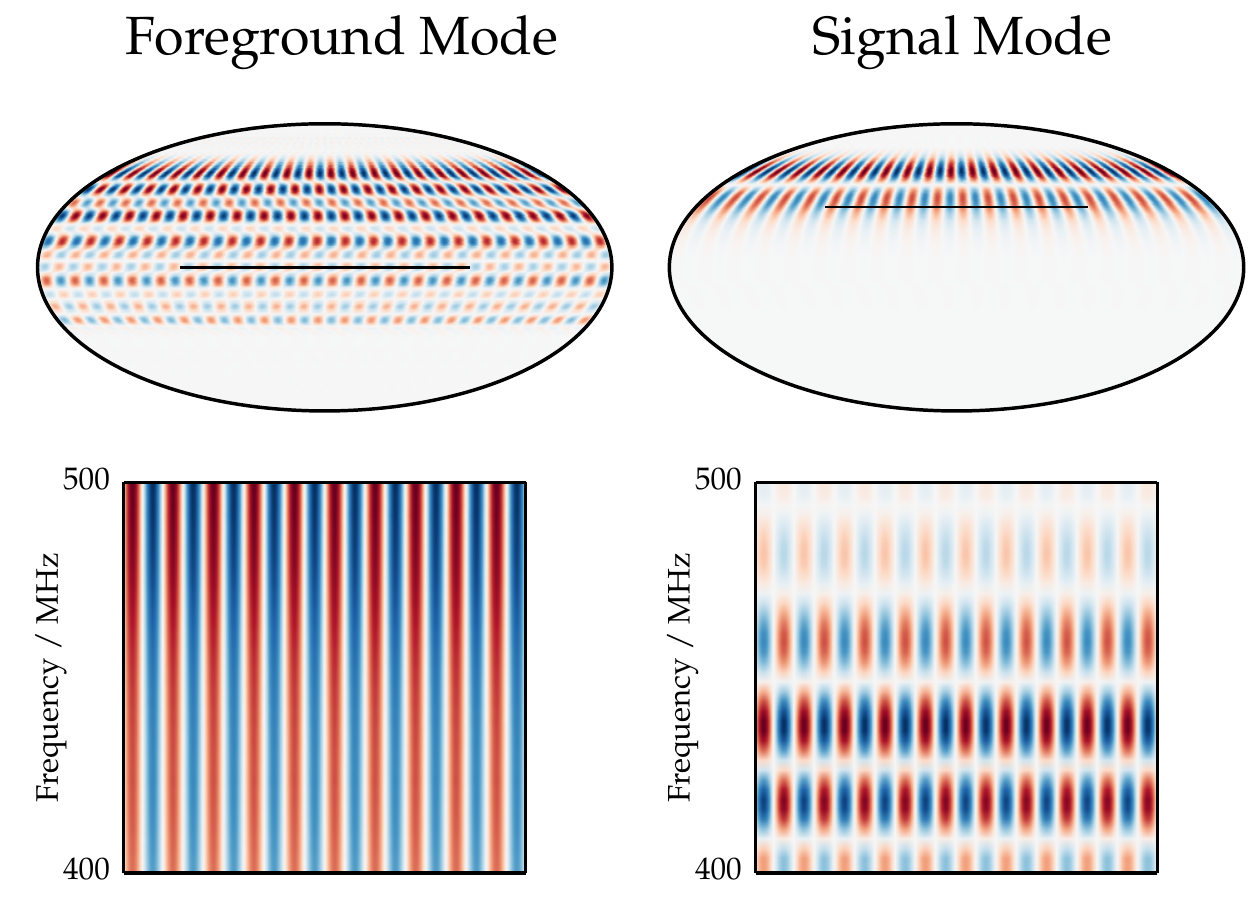}

\caption{Here we plot two KL-modes (with $m=20$) as they would look on the sky:
one of the most foreground like modes ($S/F = 4 \times 10^{-13}$); and one of
the most signal like ($S/F = 170$). Though they are derived in visibility
space, When projected back to the sky, they appear as we would expect with the
foreground mode having a smooth frequency spectrum, and the signal mode
oscillating. Modes at either end of the spectrum, like the ones plotted are
easy to interperet, this is not generally true of the intermediate modes.}
\label{fig:klvis}
\end{figure}

Transforming a visibility vector into the new basis is done by simply applying
\begin{equation}
\vvb' = \mP \vvb \; .
\end{equation}
To isolate the \tcm signal we want to select modes which contain little
foreground contamination, which can be done by picking modes with eigenvalue
(S/F power) greater than some threshold. This forms a reduced basis within
which the remaining modes have negligible contamination by foregrounds. To
project into this basis we define the matrix $\mP_s$ which contains only the
rows from $\mP$ corresponding to eigenvalues greater than the threshold $s$. In
\cref{fig:klvis} we illustrate how the signal and foreground modes appear when
projected back onto the sky.

\begin{figure}

\includegraphics[width=0.6\linewidth]{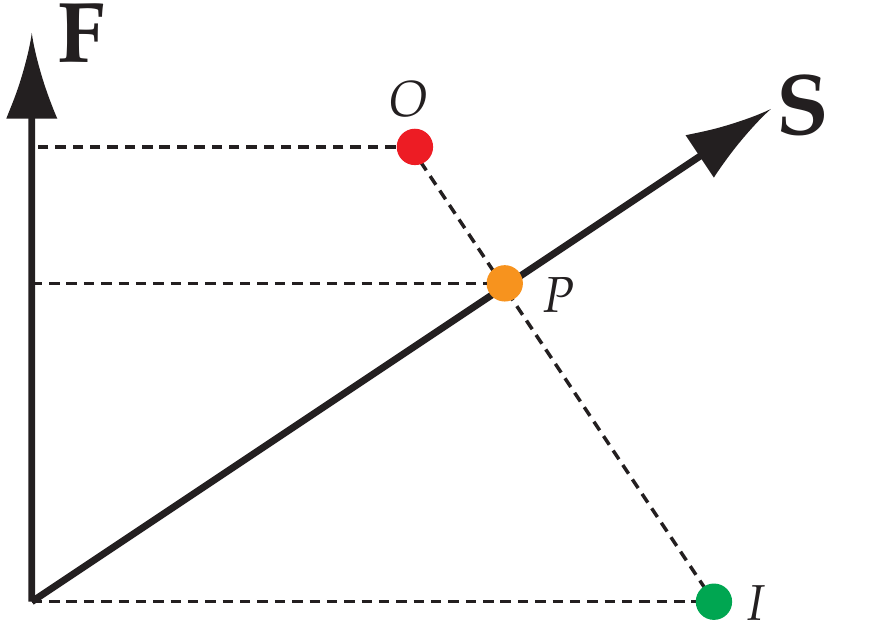}

\caption{To remove foregrounds from our data (point $O$), we separate our
space into two subspaces of foreground contaminated modes, and signal modes
(denoted by $\mathbf{F}$ and $\mathbf{S}$). These spaces are not guaranteed to
be orthogonal. Inverting with the pseudo-inverse, gives the linear combination
of signal vectors with the same amplitude, however, the resulting vector $P$
is clearly contaminated by foregrounds (as the projection onto $\mathbf{F}$ is
non-zero). The full-inverse gives point $I$, which has the same projection
onto $\mathbf{S}$, but contains no foregrounds, however, it is necessarily a
combination of both $\mathbf{S}$ and $\mathbf{F}$. }
\label{fig:invplot}
\end{figure}

For the purpose of power spectrum estimation (see next section) we will only
require forward estimators (where we project quantities into the KL-basis) and
knowing $\mP_s$ will suffice. However, for visualising our results, we want to
be able to transform back to the sky (by way of the measured visibilities).
This requires us to use an inverse to map from the truncated KL-basis back to
the visibilities. Unfortunately because the KL-modes are non-orthogonal it is
ambiguous how to project back into the higher dimensional space. One obvious
choice would be to make further use of the Moore-Penrose pseudo-inverse. This
returns a vector in the visibility space which is a linear combination of the
retained signal modes whilst preserving their projected amplitudes. However,
because the full set of modes are not orthogonal the resulting vector has a
non- zero foreground amplitude (see \cref{fig:invplot} for a visual
illustration).

A far better choice is to generate the full inverse $\mP^{-1}$ and remove
columns corresponding to the rejected modes (we denote this matrix $\mP_{-s}$).
This is equivalent to projecting into the full KL-basis, zeroing the foreground
contaminated modes, and the using the full-inverse to return the visibility
space. The distinction with the pseudo-inverse is shown in
\cref{fig:invplot}.

\begin{figure*}
\includegraphics[width=\textwidth]{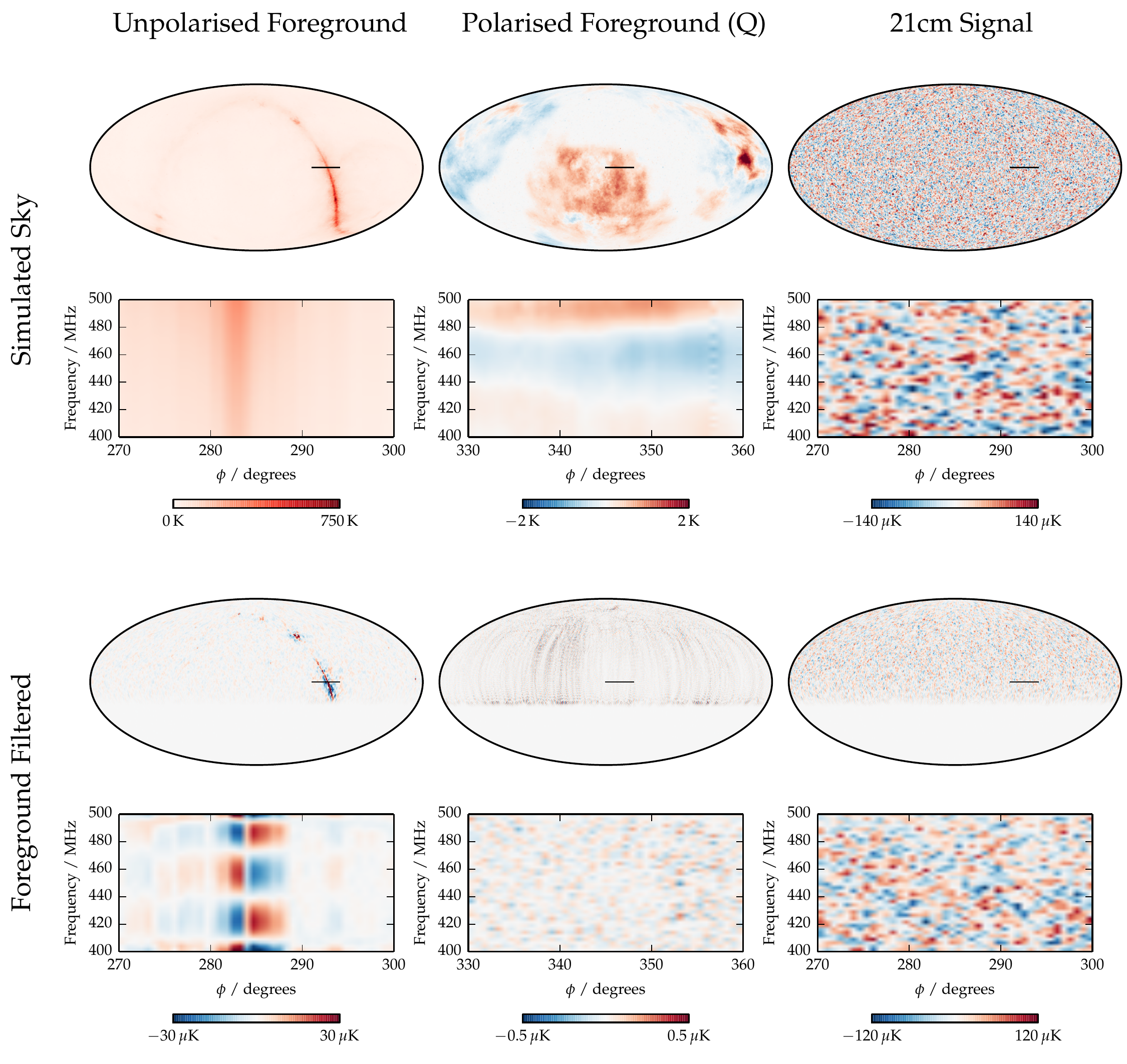}

\caption{
This plot illustrates the process of foreground removal on simulations of the
radio sky. The top row of plots show skymaps of the individual components:
unpolarised foregrounds, polarised foregrounds (showing Stokes $Q$ only), and
the \tcm signal. On the bottom row we show the maps we would make after
foreground cleaning visibilities from our example telescope. Both the
polarised and unpolarised foregrounds become substantially supressed, whereas
the \tcm signal is largely unaffected. In this example we have discarded modes
with $S/F < 10$. This leaves a clear correspondence between the original
signal simulation and the foreground subtracted signal, whilst leaving the
foreground residuals over $10$ times smaller in amplitude.}
\label{fig:foregroundremoval}
\end{figure*}

To demonstrate the foreground removal process we project separate realisations
of the total intensity foregrounds, polarised foregrounds (showing Stokes Q
only), and the \tcm signal, through the filtering process (see
\cref{fig:foregroundremoval}). We show the original simulations, and the maps
made from the foreground filtered visibilities. This illustrates how the
foreground amplitude is dramatically reduced by the process, whilst the signal
retains its overall character.

\subsection{Double-KL transform}
So far we have neglected the effects of instrumental noise. To add the
instrumental noise back in we simply transform all noise contributions into
the new basis. Writing the total noise contribution as $\mN_\text{all} = \mFb
+ \mNb$, the matrix in the truncated basis is
\begin{align}
\mN^\text{all} \rightarrow \mN^\text{all}_s & = \mP_s \lp \mFb + \mNb \rp \mP_s^\hconj \\
& = \mI + \mP_s \mN \mP_s^\hconj \; .
\end{align}
Though this transform ensures that our foreground contamination remains
minimal, as the transformed instrumental noise matrix will not remain diagonal
this gives a correlated component between all our modes. However, for further
analysis it will be particularly useful if the set of modes we use in our
calculation are uncorrelated. By making a further KL-transformation on the
foreground removed signal $\mS_s = \Lambda_s$, and total noise
$\mN^\text{all}_t$ covariance matrices, we find a new transformation matrix
$\mQ$ which maps into a basis where this is true. We will apply a further
cutoff to this,  including only modes with a signal to total noise ratio
greater than $s$ to give a transform $\mQ_t$.

For notational convenience we will write the total transformation in terms of
a single matrix $\mR = \mQ_t \mP_s$, having chosen suitable values for the two
cutoffs $s$ and $t$. Quantities in this final basis we denote with tildes, for
example a visibility mapped into this basis is $\tilde{\vv} = \mR \vvb$, and a
covariance is $\mCt = \mR \mCb \mR^\hconj$. We will denote the signal
covariance $\mSt = \mLambdat$, and the total noise covariance (including
foregrounds) as $\mNt = \mI$.


\section{Power Spectrum Estimation}
\label{sec:powerspectrum}

In cosmology we are primarily interested not in the individual structures we
see, but in their global properties. It is these statistical observations which
tell us about the fundamental nature of the Universe. The quantity we are most
interested in is the power spectrum which encodes most of the cosmological
information in its shape and evolution. In particular for \tcm intensity
mapping it allows us to measure the position of the Baryon Acoustic
Oscillations (BAOs), which in turn can shed light on the time evolution of dark
energy \cite{Blake2003}.

In order to determine the power spectrum shape we first need to parameterise
it. We choose to model the two-dimensional, real-space comoving power spectrum,
describing it as a linear summation of different basis functions
\begin{equation}
\label{eq:psbands}
P(\vk) = \sum_a p_a P_a(\vk) \; .
\end{equation}
In this paper we decompose $k$-space into bands in $k_\parallel$ and
$k_\perp$, such that each band represents a ring around the line of sight axis
in the full three dimensional $k$-space.

We can calculate the accuracy we could achieve measuring the power spectrum
using the Fisher Information Matrix, which provides a method for predicting our
ability to constrain arbitrary sets of parameters, and has become the essential
tool in cosmology for forecasting. The Fisher matrix is defined as
\begin{equation}
F_{ab} = - \la \frac{\partial^2}{\partial p_a \partial p_b} \log{\mathcal{L}(\vec{p}; \vvt)} \ra_{\vvt}
\end{equation}
where $\mathcal{L}$ is the Likelihood function and the $p_a$ are the parameters
we are trying to forecast. In the limit that we are measuring the Fisher
information for the true parameters $\vec{p} = \vec{p}_0$ that generate the
data $\vvt$, and the priors are uniform in this region, the inverse $\mF^{-1}$
gives a lower bound on the errors of any unbiased estimator (the Cram\'er-Rao
bound), and can be viewed as a forecast for the covariance of the $p_a$.


Let us specialise this to the case of estimating the power spectrum. After
projection into the foreground cleaned basis we assume that the remaining modes
follow a complex Gaussian distribution with zero mean. This assumption should
be reasonable provided we have successfully removed the modes containing any
significant foreground contribution---it is these modes which contain the most
non-Gaussian contributions. In this case the Fisher Information matrix of a
single $m$-mode for a set of parameters $p_a$ is
\begin{equation}
\label{eq:fisher_gaussian}
F_{ab}^\brsc{m} = \Tr{\ls \mCt_a \mCt^{-1} \mCt_b \mCt^{-1} \rs} \; .
\end{equation}
where $\mCt_a$ is the linear response of the data covariance to a change in
$p_a$, that is
\begin{equation}
\mCt_a = \frac{\partial \mCt}{\partial p_a} = \frac{\partial}{\partial p_a} \la \vvt \vvt^\hconj \ra \; .
\end{equation}
For power spectrum forecasting, the $p_a$ are the amplitudes of our power
spectrum bands (see \cref{eq:psbands}). To calculate the response $\mCt_a$ we
need to project the band functions $P_a(\vk)$ into the cleaned basis. First,
the spatial representation $P_a(\vk)$ must be turned into a multi-frequency
angular power spectrum $C_l(\nu, \nu') = \la a_{lm}(\nu) a^*_{lm}(\nu') \ra$.
We do using a simple linear flat-sky prescription which both includes the
effects of redshift distortion and structure growth (see \cref{app:models}). We
denote the matrix representation of the angular power spectrum basis function
as $\mC_a$. This must be projected into the KL-basis
\begin{equation}
\label{eq:ca_proj}
  \mCt_{a} = \mR \mB \mC_a \mB^\hconj \mR^\hconj \; .
\end{equation}
In practice explicitly calculating the $\mCt_a$ this way is computationally
very expensive. We will discuss a fast Monte-Carlo alternative for calculating
the Fisher matrix later in this section. In the constructed eigenbasis $\mCt =
\mLambdat + \mI$ is exactly diagonal, however, $\mCt_a$ can have off-diagonal
elements.

As there is no coupling between them, the total Fisher Information for the
whole dataset is simply the sum over the individual $m$-modes
\begin{equation}
\label{eq:fisher_combined}
F_{ab} = \sum_m F_{ab}^\brsc{m} \; .
\end{equation}

The Fisher matrix gives us the ability to forecast how well we can possibly
measure the power spectrum, but it does not tell us how to go about estimating
that power spectrum. We will use the quadratic power spectrum estimator of
\cite{Tegmark1997,Bond1998}. This is an \emph{optimal} estimator in that it
achieves the Cram\'er-Rao bound giving it the lowest possible variance. We will
give an overview of this estimator below, though we encourage the reader to
look at the original papers for more detail.

As our dataset is made up of a large number of independent $m$-modes, for
simplicity we will start with the power spectrum estimator for the whole
dataset, and then break it down into individual modes which can be calculated
simply.

For notational simplicity, it is most convenient to start with a related
estimator
\begin{equation}
\label{eq:qhat_def}
\hat{q}_a = \vvt^\hconj \mE_a \vvt \; ,
\end{equation}
where the quantities on the right hand side include all $m$'s. This forms a
weighted combination of all the quadratic pairs $\vvt \vvt^\hconj$. Our actual
power spectrum estimator is built out of linear combinations of the
$q$-estimator.
\begin{equation}
\hat{p}_a = \sum_b M_{ab} \lp \hat{q}_b - b_b \rp \; .
\end{equation}
In this $b_b$ subtracts the additive bias from the instrumental and foreground
noise, and the mixing matrix $M_{ab}$ takes linear combinations such that
$\hat{p}_a$ is related to the actual power spectrum. Our estimator will have
minimum variance with the choice
\begin{equation}
\label{eq:mE_def}
\mE_a = \mCt^{-1} \mCt_a \mCt^{-1} \; ,
\end{equation}
where again these matrices include all $m$'s. For a detailed derivation of
this weighting, see \cite{Tegmark1997}. Requiring $\hat{p}_a$ to be an
unbiased estimator of the power spectrum we can determine the noise bias term
\begin{equation}
b_a = \Tr{\mE_a \mNt} \; .
\end{equation}

Our remaining choice is that of the mixing matrix $M_{ab}$ which gives the
exact link between our estimator, and the the `true' power spectrum. In
particular, we care about the expectation of the estimator
\begin{equation}
\label{eq:phat_exp}
\la \hat{p}_a \ra = \sum_b W_{ab}\, p_b \; ,
\end{equation}
which we have written in terms of a window function $W_{ab}$ which mixes the
power spectrum bands. Using \eqref{eq:qhat_def} and \eqref{eq:mE_def} we find
that
\begin{align}
\la \hat{q}_a - b_a \ra &= \Tr{\ls \mE_a (\mCt - \mNt) \rs} \\
& = \sum_b F_{ab} \, p_b \,
\end{align}
and combining this with \eqref{eq:phat_exp}, gives the window function as
\begin{equation}
W_{ab} = \sum_c M_{ac} F_{cb} \; .
\end{equation}
We fix the normalisation by requiring that $\sum_b W_{ab} = 1$. Our choice of
the mixing matrix $M_{ab}$ also affects the covariance of the estimator,
giving
\begin{equation}
\Cov(\hat{p}_a, \hat{p}_b) = \sum_{c d} M_{ac} M_{bd} F_{ab} \;,
\end{equation}
where we have used the fact that $\Cov(\hat{q}_a, \hat{q}_b) = F_{ab}$.

There are three common choices for the mixing matrix $M_{ab}$
\cite{Padmanabhan2003}:
\begin{description}
\item[Unwindowed]
Choosing the window function to be the identity means that $\la \hat{p}_a \ra
= p_a$. This corresponds to $M_{ab} = F^{-1}_{ab}$. This is the most natural
choice, however it gives highly correlated errors bars.

\item[Uncorrelated] To make the estimator covariance, we choose $M_{ab} = \ls
\sum_b F^{1/2}_{ab} \rs^{-1} F^{-1/2}_{ab}$. This leads to uncorrelated
estimates, but leads to mildly spread window functions \cite{Dillon2012}.

\item[Minimum Variance] The minimum variance estimator requires that the
mixing matrix is diagonal $M_{ab} = \ls \sum_c F_{ac} \rs^{-1}$, and gives
window functions with moderate spread.
\end{description}
We are generally interested in the Unwindowed estimator and we will use this in
our forecasts, however, for convergence reasons that we discuss later, we will
also use the Minimum Variance estimator when estimating power spectrum biases.

Naive calculation of this estimator is problematic because of the large
dimensionality of the data. However, we can trivially exploit the independence
of the individual $m$-modes to simplify this calculation. Noting that the
covariance matrices in \eqref{eq:mE_def} are block diagonal in $m$ because they
are statistically independent, the weight matrix $\mE_a$ is also block
diagonal. This means we can rewrite the q-estimator as a sum of seperate
estimators for each $m$
\begin{equation}
\hat{q}_a = \sum_m \hat{q}_a^\brsc{m} \; ,
\end{equation}
with
\begin{equation}
\hat{q}_a^\brsc{m} = \vvt_m^\hconj \mE_a^\brsc{m} \vvt_m \; ,
\end{equation}
where the $\mE_a^\brsc{m}$ are the diagonal blocks of $\mE_a$, and $\vvt_m$ is
the data for each $m$-mode. Similarly we will also break up the bias terms into contributions from each $m$. The total bias
\begin{equation}
b_a = \sum_a b_a^\brsc{m}
\end{equation}
where the individual
\begin{equation}
b_a^\brsc{m} = \Tr{\mE_a^\brsc{m} \mNt^\brsc{m}} \; .
\end{equation}

Unfortunately exact calculations of the Fisher matrix $F^\brsc{m}_{ab}$ and the
bias $b^\brsc{m}_a$ are computationally difficult. While many aspects of the
calculation can be simplified by the fact that $\mCt$ is diagonal, the need to
explicitly construct covariances of $\mCt_a$ and $\mNt$ in the KL-basis is
still prohibitive. To avoid this, we follow \cite{Padmanabhan2003,Dillon2012}
and construct a Monte-Carlo scheme to evaluate the Fisher matrix.

The key to this scheme is that evaluating the $q$-estimator is quick as we do
not need to explicitly construct any large matrices. This is achieved by
constructing the intermediate vector
\begin{equation}
\vw =  \mBb^\hconj \mR^\hconj \mCt^{-1} \vvt \; ,
\end{equation}
which can be efficiently evaluated from right to left. The $q$-estimator is
then
\begin{equation}
\hat{q}_a^\brsc{m} = \vw^\hconj \mC_a \vw  \; .
\end{equation}
As $\mC_a$ is block diagonal in $l$ this can be quickly evaluated. To estimate
the Fisher matrix we draw many random realisations of our dataset to which we
apply the $q$-estimator. Then, noting that the covariance of $\hat{q}_a$ is
\begin{equation}
\Cov(\hat{q}^\brsc{m}_a, \hat{q}^\brsc{m}_b) = F^\brsc{m}_{ab}
\end{equation}
we can evaluate the sample covariance of our $q$-samples to form an estimate
of the Fisher matrix. The estimate the bias term we use the fact that
\begin{equation}
b_a^\brsc{m} = \Bigl\langle \vnt^\hconj \mE_a \vnt \Bigr\rangle = \Bigl\langle q_a^\brsc{m} \Bigr\rangle_{\mNt} \; ,
\end{equation}
and take the average of the $q$-estimator under random realisations of the
noise.

This Monte-Carlo scheme converges rapidly enough that it is effective for
forecasting. Unavoidably there will be small off diagonal terms in the Fisher
matrix which do not converge exactly, and these errors can become amplified
when taking powers to construct the mixing matrix $M_{ab}$. These errors remain
small enough that they are not apparent when performing power spectrum
estimation on data close to the fiducial model, and in most cases this
Monte-Carlo technique is still sufficient. However, for data significantly
biased from the fiducial model (by $10^2$--$10^3$ times the estimator error
bar) these errors can add spurious noise to the estimator. This is particularly
acute when using the Unwindowed estimator which requires the inverse of the
Fisher matrix. However, using the Minimum Variance estimator, which does not
require us to calculate any powers of the Fisher matrix, alleviates this
problem. This is the route we take when dealing with the  biased data we
will find in \cref{sec:uncertain}.


\section{Discussion}
\label{sec:discussion}

\subsection{Polarised Foreground Removal}
\noindent
Foreground cleaning inevitably throws away measured information about the sky,
and is guaranteed to reduce our sensitivity to the \tcm signal we are seeking.
As our primary interest is to measure the \tcm power spectrum it is vital that
we understand how foreground cleaning methods affect our power spectrum errors.
Over the previous sections we have developed the tools to tackles this: in
\cref{sec:formalism} we saw how the $m$-mode formalism gives us a simple and
efficient description of the measurement process; \cref{sec:kltransform}
developed an effective foreground cleaning method based on the KL-transform
that allows us to easily tracks the statistics of our data through the
cleaning; and in the previous section (\cref{sec:powerspectrum}) we constructed
an optimal estimator for the power spectrum, and forecast its errors using the
Fisher matrix. Here, we combine these to forecast the performance of our
example telescope in the presence of foregrounds.

\begin{figure}
\includegraphics[width=\linewidth]{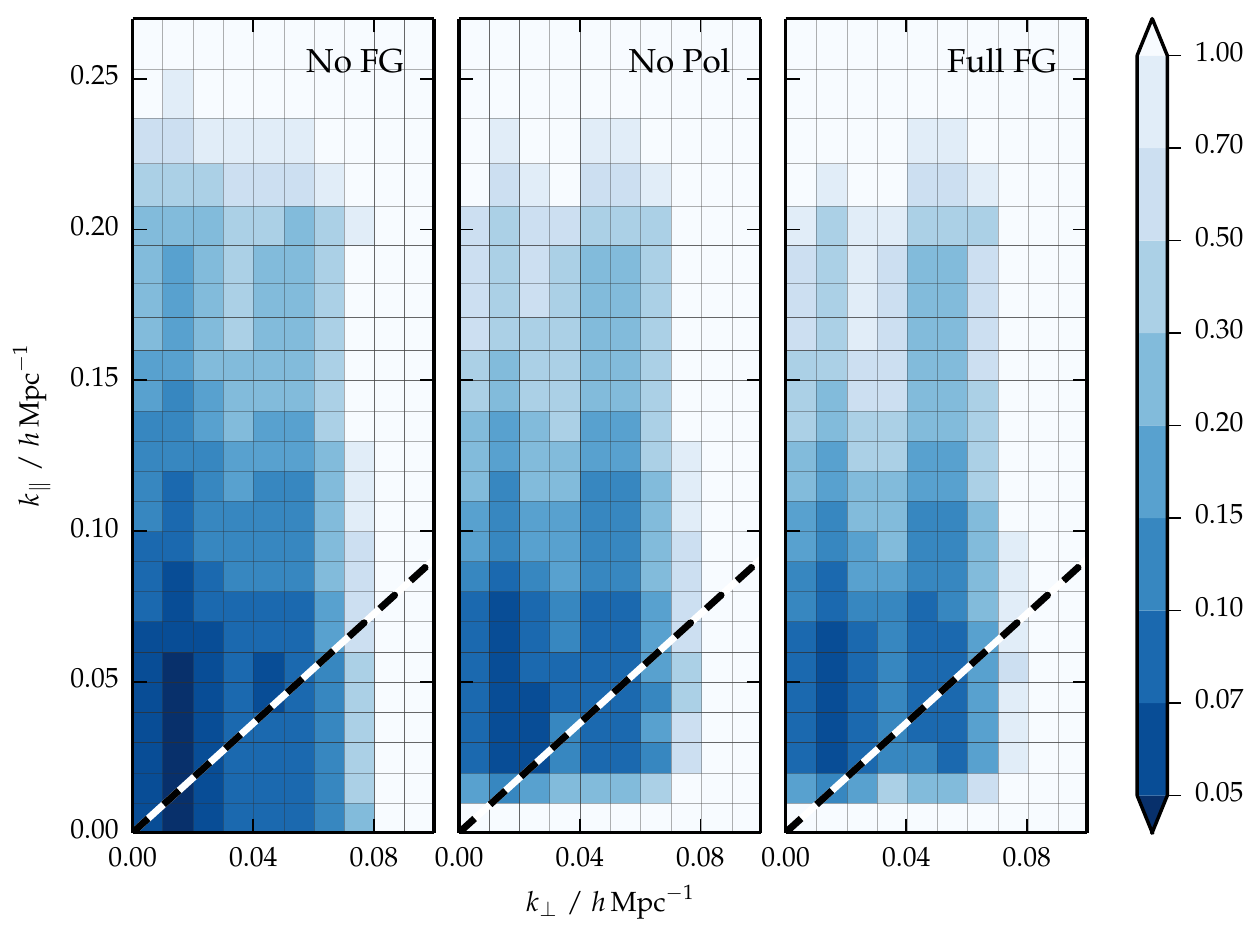}

\caption{Forecast errors on the power spectrum as a fraction of its fiducial
value for the \SIrange{400}{500}{\mega\hertz} band. The three panels show the
predicted errors without foregrounds (left), with unpolarised foregrounds
(centre), and with fully polarised foregrounds (right). The dashed line
indicated the predicted bound of the `foreground wedge', showing that with
perfect knowledge of our instrument foregrounds can be successfully cleaned
well into this region.}

\label{fig:pscomp_fg}

\end{figure}

In \cref{fig:pscomp_fg} we show the power spectrum errors for observations of
the \SIrange{400}{500}{\mega\hertz} band with our example telescope. We forecast three distinct sets of
foregrounds: \emph{no} foregrounds; completely \emph{unpolarised} foregrounds;
and \emph{partially polarised}. We use values of the foreground amplitudes and
spectral correlation that are representative of those in our galaxy, these
models are described in detail in \cref{app:models}. In particular the latter
includes the effects of Faraday rotation, especially emission from a range of
Faraday depths within our galaxy, that produces significant spectral structure
in the polarised emission.

Clearly the dominant effect of foreground removal in both cases is that we
become insensitive to power at low $k_\parallel$, with a slight increase in
the errors across $k$-space. This is in line with our expectations that the
foregrounds contaminate the large scale frequency modes corresponding to small
$k_\parallel$, though we discuss this how this relates to the \emph{foreground
wedge} of \cite{Morales2012,Parsons2012} later.

Polarised foregrounds are removed primarily by the action of the SVD filter
described in \cref{sec:svd}. This leads to only a slight worsening of the
errors compared to the case of unpolarised foregrounds only. One concern could
be that the SVD filter does not discriminate between polarised modes on the
basis of the magnitude of their contamination (as would be done by a KL-based
filter), it removes them all. This approach is not perfectly optimal, and could
be improved by allowing all polarisation modes to propagate through and let the
KL-filter determine which to remove. In tests on smaller examples, this
approach yields no significant improvement, but due to computational
limitations can not be demonstrated on the example in this work.

In all the cases illustrated in \cref{fig:pscomp_fg} there are clear peaks in
the sensitivity in the $k_\perp$ direction that correspond to those seen in
\cref{fig:fisher_sensitivity}, and a rapid drop-off as we approach the limit
of resolution limit of the telescope. Additionally at low $k_\perp$ we can see
there is a reduction in sensitivity caused by the sample variance of the small
number of large scales angular modes.

\subsection{Foreground Wedge}

Previous studies of the performance of \tcm experiments in the face of large
astrophysical foregrounds have found the bulk of the contamination to lie in a
wedge shaped region of $k_\parallel < \beta k_\perp$ (for an experiment
dependent constant $\beta$), termed the \emph{foreground wedge}
\cite{Datta2010,Parsons2012,Morales2012}. In these studies, the complement of
this region remains largely free of contamination, and is thought to provide
the best chance for observing cosmological \tcm radiation (in the context of
Epoch of Reionisation observation this region is called the \emph{EoR
Window}).

Important progress has been made in recent years understanding the source of
this contamination \cite{Parsons2012,Morales2012}: spectrally smooth radio
emission is observed at a delay which depends on the baseline length, and
distance of the emission from the phase centre, the phase rotation with
frequency from this delay appears like fluctuations along the line of sight.
This argument leads us to predict that spectrally smooth sources contribute
power within a region
\begin{equation}
k_\parallel < \Delta\theta \lp \chi(z) \frac{H(z)}{c (1 + z)} \rp k_\perp \; ,
\end{equation}
where $\chi(z)$ is the comoving distance to redshift $z$, and $\Delta\theta$
is the maximum observable distance from the beam centre.

In \cref{fig:pscomp_fg} we mark the boundary of the foreground wedge for our
example telecope. While foreground removal makes us insensitive to small
$k_\parallel$ there is no discernible variation of this with $k_\perp$, and we
can observe well into the `foreground wedge'. Clearly there is no fundamental
loss of information about the entire wedge. Though the distinction between the
information lost here, and the whole wedge is small for our example, for a
larger telescope with higher angular resolution the difference will be
significant.

As pointed out in \cite{Liu2012} we expect the foregrounds along each line of
sight to be described by only a small number of eigenmodes (with those beyond
five contributing less than $10^{-10}$ in power). Though mode-mixing may make
these modes appear to contribute power throughout the foreground wedge,
fundamentally there are only a small number of them. The KL-transform projects
these eigenmodes forward into the data-basis while keeping track of how their
angular structure correlate different baselines. If our knowledge of the
telescope is perfect, we can use these modes to exactly project out the large
foreground contributions to the data.

If our knowledge of the telescope is not perfect as in our forecasts (e.g.
\cref{fig:pscomp_fg}), we cannot perfectly remove the foregrounds. We
investigate this in the following section.

\section{An Uncertain World}
\label{sec:uncertain}

So far we have demonstrated that the \tcm signal can be separated from the
astrophysical foregrounds in a way which does not distort our measurement of
the underlying power spectrum. This assumed an ideal instrument about which
our knowledge was perfect in every sense, conditions that a real telescope
will not meet. There are many sources of non-ideality --- primary beam
response, amplifier gains, cable delays and noise temperatures are just a few
--- each of which could distort our measurements. We can divide these
non-idealities into two classes:
\begin{itemize}

\item \emph{Known deviations} from the design can be incorporated into our
analysis to keep it unbiased and optimal, though our ultimate sensitivity may
change relative to the design.

\item \emph{Unknown deviations} from our best model of the instrument cannot be
corrected and will lead to bias from both foreground leakage and using a biased
power spectrum estimator.

\end{itemize}
The second class of deviations is the most serious, and so for these effects we
would like to know how large our uncertainty can be before it matters, or more
precisely before it is significant compared to the statistical errors.

As our ability to separate signal and foregrounds requires detailed knowledge
of our instrument, we can form a naive expectation of the allowed uncertainty
from the dynamic range between signal and foregrounds. In the smooth frequency
modes where foregrounds dominate, they are around $10^5$ times brighter than
the \tcm (\SI{10}{\kelvin} versus \SI{0.1}{\milli\kelvin}), and so we expect
that knowing our instrumental gains and beam shapes to $10^{-5}$ accuracy
should be sufficient.

In this section we aim to test two particular forms of uncertainty that we can
parametrise simply in our model telescope to see if the requirements are as
stringent as $10^{-5}$. Our approach is to assume that our example telescope
represents our best knowledge about the state of the system, which we use to
generate our foreground cleaning filter and our power spectrum estimator. We
then generate a \emph{corrupted timestream} corresponding to the observations
the true telescope would make. By analysing this timestream with the filters
generated for the example telescope we can see at what point imperfect
knowledge leads to significant power spectrum biasing.

\subsection{Gain Fluctuations}

A receiver system turns the input antenna voltage into a signal which can be
measured and correlated. In the process of doing this a complex gain may be
applied, and while this can be corrected for, this generally leaves unknown
residuals in the data. This gain residual is unique to each feed and may be
time and frequency dependent.

We model gain fluctuations on a feed by feed basis, as a complex perturbation
around a nominal gain of unity. The perturbed feed input is 
\begin{equation}
F_i' = (1 + \Delta{g}_i) F_i \,
\end{equation}
where the perturbation $\Delta{g}$ is a complex Gaussian random variable with
variance $\la \Delta{g}_i \Delta{g}_i^* \ra = \sigma_g^2$. These combine to
give corrupted visibilities
\begin{equation}
V_{ij}' = (1 + \Delta{g}_i) (1 + \Delta{g}_j^*) V_{ij} \; .
\end{equation}

In our model we do not allow the gain to fluctuate in frequency, enforcing each
antennas gain to be frequency independent. However, we do allow the gains to
fluctuate in time, assuming that each \SI{60}{\second} sample has a separate
uncorrelated gain residual. Over the two years of integration, the errors on
each co-added sample are reduced by a factor of $\sqrt{733}$.

\begin{figure}
\includegraphics[width=\linewidth]{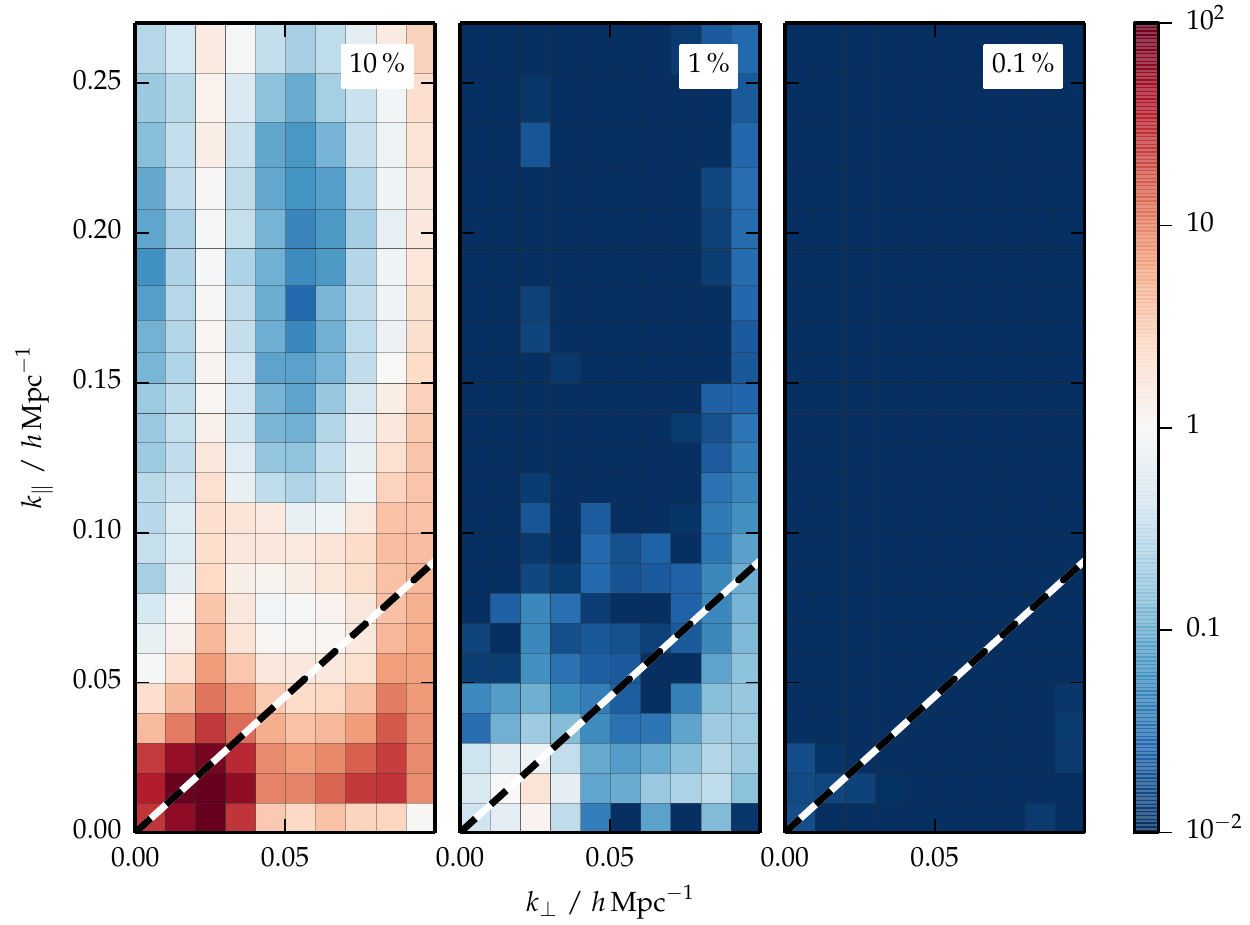}

\caption{Biasing of the power spectrum from complex gain perturbations with
amplitude $\sigma_g = 10\%$, $1\%$ and $0.1\%$, again for observations of
\SIrange{400}{500}{\mega\hertz}. The bias is given as a fraction of the
statistical error. Regions where this ratio is less than one (shown in blue)
indicate where the systematic errors are sub-dominant compared to the
statistical errors. Again we indicate the foreground wedge with a dashed line,
however in this case we note that most of the bias lies within this region.}

\label{fig:bias_gain}

\end{figure}

We start with the base timestream to which we have added random gain
fluctuations with $\sigma_g = 10\%$, $1\%$ and $0.1\%$ in each
\SI{60}{\second} period. These time streams are then analysed with the
fiducial analysis products that assume no gain fluctuations. In
\cref{fig:bias_gain} we show the power spectrum biases corresponding to each
level of gain fluctuation. We have used the Minimum Variance estimator
discussed in \cref{sec:powerspectrum}, the results appear similar if we use
the Unwindowed estimator (albeit noisier). The bias, which is caused by
foreground leakage from the imperfect calibration, is mostly located within
the foreground wedge. This is inline with our expectation from
\cite{Morales2012,Parsons2012} which indicate that leakage from imperfect
foreground cleaning will be concentrated in this region. However, there are
significant discrepancies from this picture that seem to be related to the
array geometry (such as the line $k_\perp = 0.03\,\ihMpc$), that may require
more detailed study to understand intuitively \cite{Hazelton2013}.

We can see that the bias becomes negligible for errors residuals of around $1
\%$. Over the course of the two years observation this corresponds to a
tolerance on gain fluctuations of $\sim \num{2e-4}$ for each synthetic beam
($\sim \SI{1}{\degree}$). This required tolerance is significantly less than
the $10^{-5}$ naively expected. This difference is due to the fact that we
repeatedly measure the same sky because our array is highly redundant (with
typical redundancies of $\sim 30$) allowing us to average down the affect of
gain fluctuations, reducing the precision required on an individual baseline.

This level of precision should be achievable with techniques such as redundant
baseline calibration \cite{Liu2010b}. Our analysis assumes that the residuals
are Gaussian and independent in time, such that they quickly average down with
repeated measurements. In practice there may be a component of the residuals from $1/f$ noise
with large correlation times which make this assessment more difficult.  We leave investigation 
of such effects for future studies.

\subsection{Unknown Primary Beam}

One of the key inputs to our analysis is an accurate model of each feeds
primary beam. In particular we need the electric field response at each
position on the sky, given by the quantity $A_a(\vnhat)$. Generally this
quantity can only be determined by calibrating from observations of the sky
(for instance by holography). As this process is challenging and
time-consuming, we would like to know how precise the calibration must be.

\begin{figure}

\includegraphics[width=\linewidth]{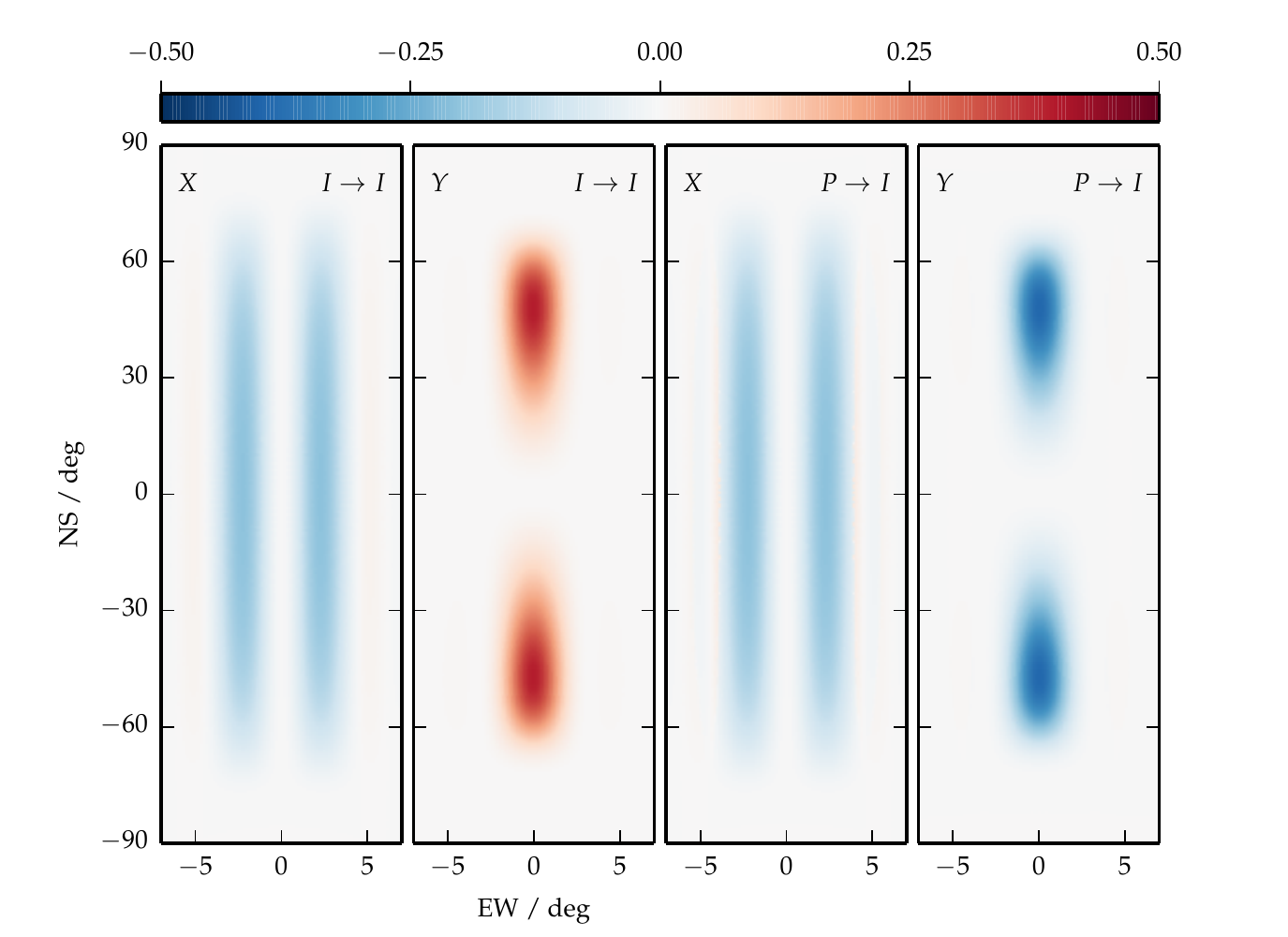}

\caption{The response of the primary beam to fractional changes in the $X$ and
$Y$ dipoles E-plane widths. Similar to \cref{fig:beam} we illustrate the
transfer from the total intensity and polarised sky, into an instrumental
Stokes I combination, however, here we show the derivative with respect to
changes in the E-plane width of the X and Y feeds. The first two plots show
the change of the total intensity response with changes in the E-plane of the
X and Y dipoles; the second two plots show the changes in the polarisation
response, again corresponding to changes in the X and Y feeds. For instance a
$1 \%$ change in each dipoles width changes each response by $1 \%$ of the
corresponding plot (to first order).}
\label{fig:beam_pert}
\end{figure}

Here, we use the parametrisation of the primary beam given in
\cref{sec:beammodel}. We use the fiducial model of the dipole's beam, $\theta_H
= 2\pi/3$, $\theta_E = 0.7 \theta_H$ (this is the same as the example used
throughout). However, we will perturb the E-plane widths of each antenna
around the fiducial model by an amount $\Delta\theta_E^i$. Increasing
$\theta_E$ has the effect of making the primary beam of the $X$-feed slightly
narrower, and the $Y$-feed longer (decreasing it does the opposite). It also
reduces the difference in response between the $X$ and $Y$ feeds, reducing the
expected amount of polarisation leakage. This is demonstrated in
\cref{fig:beam_pert} where we show the effect on the Stokes I and polarised
response to changes in $\theta_E$ to the $X$ and $Y$ feeds. In particular we
show the derivatives of $R_{I \rightarrow I}$ (\cref{eq:rII}) and $R_{P
\rightarrow I}$ (\cref{eq:rPI}) with respect to $\theta_E^X$ and $\theta_E^Y$.

\begin{figure}
\includegraphics[width=\linewidth]{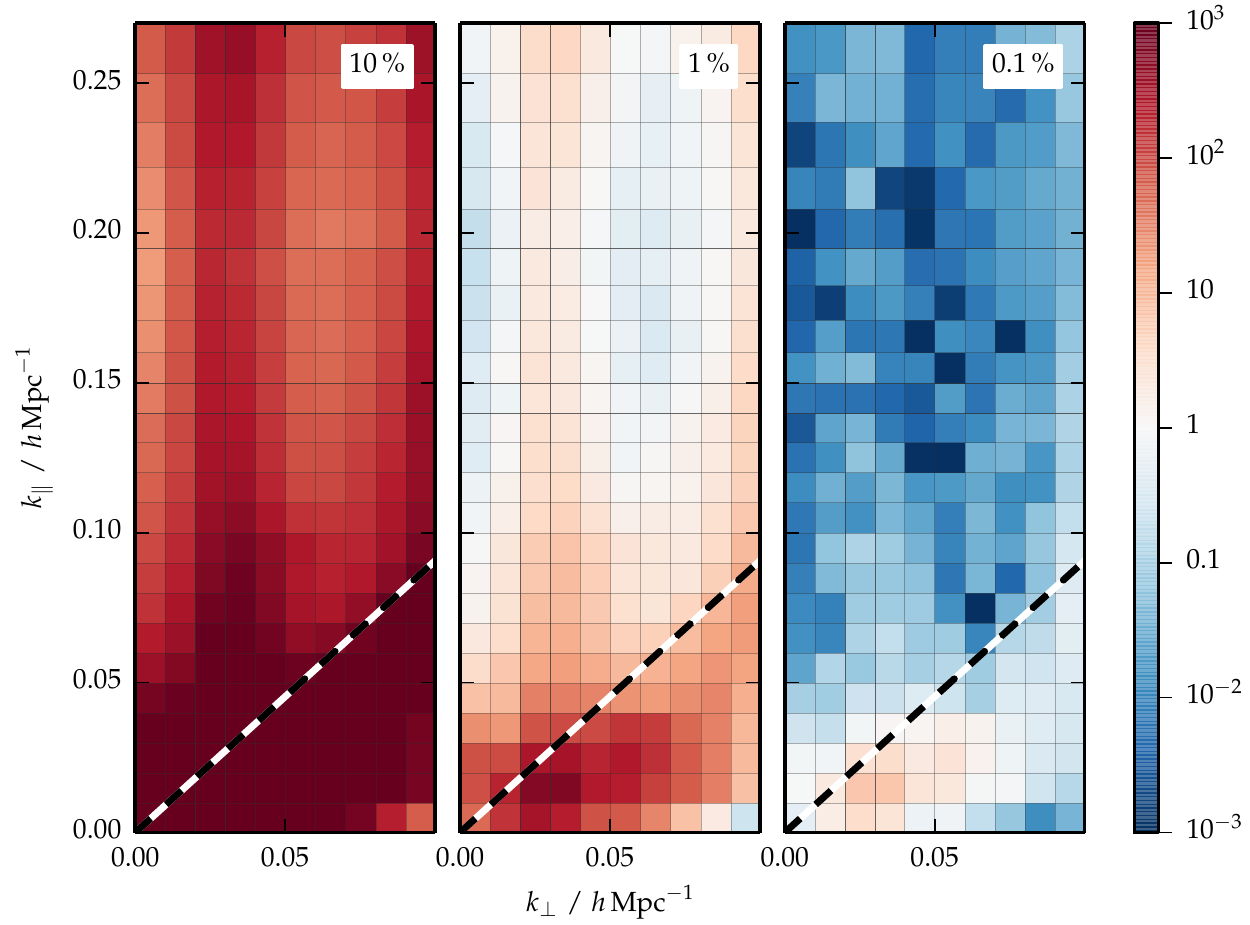}

\caption{Power spectrum biasing for $10 \%$, $1 \%$ and $0.1 \%$ shifts from
the fiducial E-plane width. These biases are given in units of $\sigma$ for
each band, values greater than one indicate where this systematic error
dominates the statistical error. These are the biases of the Minimum Variance
estimator (so as to avoid issues with the power spectrum deconvolution). Here
we can see that unknown fluctuations in the beam width of more than $0.1 \%$
give rise to significant power spectrum biases.}

\label{fig:bias_squint}
\end{figure}

To calculate the changes to the data we need to propagate these primary beam
changes through to the Beam Transfer matrices. At linear order in the
$\Delta\theta_E^i$ the perturbed Beam Transfer functions are
\begin{equation}
\label{eq:btexpand}
B_{ij}^X = B_{ij}^X + \frac{d B_{ij}^X}{d\theta_E^i} \Delta\theta_E^i + \frac{d B_{ij}^X}{d\theta_E^j} \Delta\theta_E^j
\end{equation}
where the derivatives are related to the primary beam derivatives by
\begin{multline}
\label{eq:btderiv}
\frac{d B_{ij}^X}{d\theta_E^k} = - \frac{d \ln{\Omega_{ij}}}{d\theta_E^k} B_{ij}^X \\
+ \frac{2}{\Omega_{ij}}\ls \frac{d A_i^a}{d\theta_E^k} A_j^{b *} + A_i^a \frac{d A_j^{b *}}{d\theta_E^k}\rs \calP^X_{ab} \:e^{2 \pi i \vnhat \cdot \vu_{ij}} \; .
\end{multline}
The derivative of the composite beam solid angle is
\begin{equation}
\frac{d \ln{\Omega_{ij}}}{d\theta_E^i} = \frac{d \ln{\Omega^*_{ji}}}{d\theta_E^i} = \frac{1}{2 \Omega_i} \int \dhn \frac{d A_i^a}{d\theta_E^k} A_j^{b *} \calP^I_{ab} \; .
\end{equation}
By treating the primary beam derivatives $d A_i^a / d\theta_E^k$ as a modified
beam, we can use \cref{eq:btderiv} to calculate timestreams for the beam
perturbed Beam Transfers. We then use \eqref{eq:btexpand} to apply the effects
of arbitrary combinations of perturbations to $\theta_E$ for each
antenna.

We draw a set of Gaussian distributed values for the width of each feed,
$\theta_E^i$. We vary standard deviation ($10 \%$, $1 \%$ and $0.1 \%$ of the
fiducial model) and use them to generate synthetic data with perturbed beam
widths. We propagate the analysis of these corrupted timestreams all the way
through to the power spectrum, assuming the fiducial configuration. In
\cref{fig:bias_squint} we show the results for the Minimum Variance estimator.
Again we see that the bias is mostly concentrated in the foreground wedge
region. The bias can be significant (compared to statistical errors) if our
beam knowledge is imperfect, though it has mostly disappeared in the case
where we know the beam width to $0.1 \%$.

This analysis suggests that if the beam width were the only varying parameter,
in the absence of other bias mitigating techniques, we would need to measure
it to $\sim 10^{-3}$ accuracy. However, as the beam derivative is typically of
order $0.1$ (see \cref{fig:beam_pert}), this can be seen as a precision of
around $10^{-4}$ on the beam itself, similar to the gain fluctuations, and
still a lower precision than our expectation of around $10^{-5}$. We can
attribute this to the fact that our power spectrum estimation is dependent on
a complicated combination of all the primary beams, and this averages down the
fluctuations in the same manner as we expect for the gain fluctuations.

Clearly a realistic description of the beam must contain much more than a
simple beam width, but this indicates the accuracy to which we must strive to
map the primary beam of each feed. This level of precision will be
challenging, though not unprecedented, with similar accuracies achieved by
holographic means
\cite{ATAcal}.

\section{Full bandwidth forecasts}
\label{sec:fullbandwidth}

Experiments such as CHIME are targeted at measuring the evolution of dark
energy over a large range of redshift. As an example application of this
method we show in this Section forecasts for the example cylinder telescope
(similar in size to the CHIME Pathfinder but smaller than full CHIME) across a
full octave in bandwidth of \SIrange{400}{800}{\mega\hertz}, corresponding to
a redshift range of $z \approx 0.8$--$2.6$. This is broken up into four
\SI{100}{\mega\hertz} sub-bands to illustrate the changes with frequency.

\begin{figure}
\includegraphics[width=\linewidth]{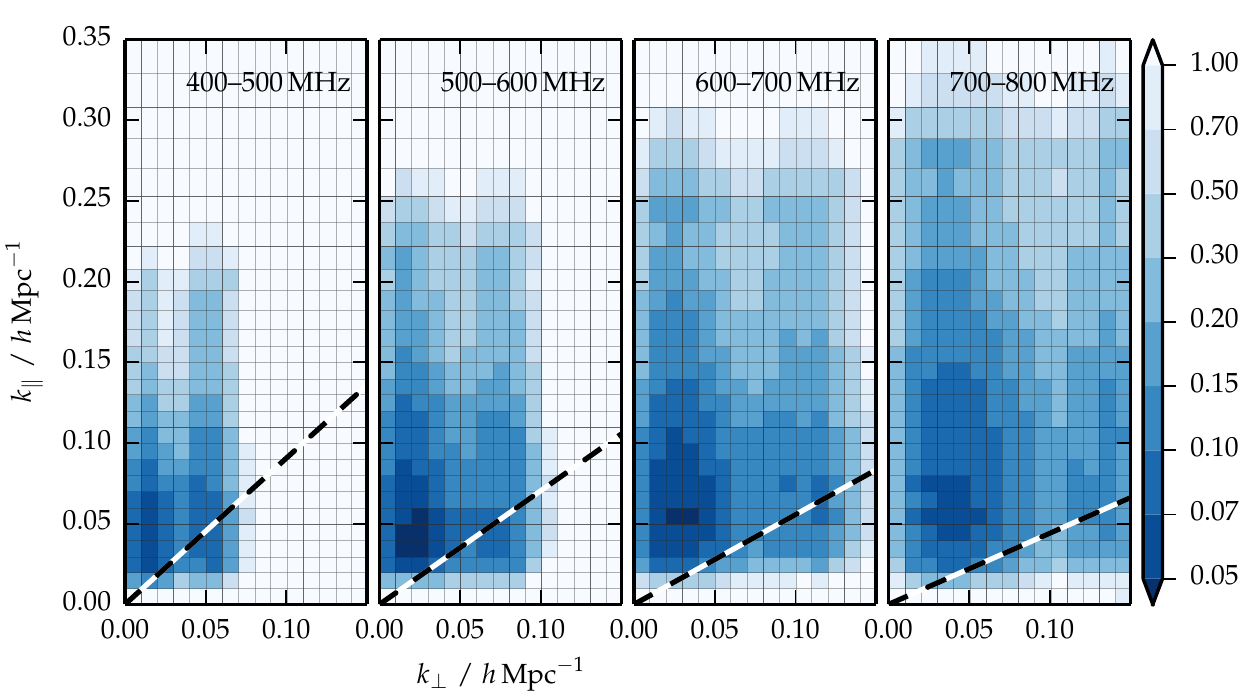}

\caption{The power spectrum sensitivity for different frequency bands between
\SI{400}{\mega\hertz} and \SI{800}{\mega\hertz}. This clearly illustrates the
increasing angular resolution as we move to higher frequencies. Again, the red
dashed lines indicate the location of the foreground wedge.}

\label{fig:ps_allbands}
\end{figure}

In \cref{fig:ps_allbands} we show the power spectrum forecasts for each of the
four \SI{100}{\mega\hertz} sub-bands. This clearly illustrates the increase in
sensitivity as we move to higher frequency, particularly at large $k_\perp$
where the increased angular resolution combines with the decreased observation
distance to dramatically increase the spatial resolution. There is an
additional boost at large $k_\parallel$ where the constant frequency
corresponds to a decreasing line of sight distance. We can also see how the
double peaked structure in sensitivity (discussed in \cref{sec:discussion})
changes with frequency, with the peaks moving outwards and broadening as
expected from the increasing resolution. However, the drop-off at small
$k_\perp$ barely increases in size as it comes from the contribution of sample
variance which does not change with the increased angular resolution (it does
shift slightly because a fixed angular scale maps a smaller spatial scale at
higher frequency).

The effect of foreground cleaning is similar across all bands, with it
removing sensitivity for $k_\parallel < 0.02 \ihMpc$. We don't expect the
number of modes used to describe the foregrounds along a particular line of
sight to vary significantly with the small shifts in frequencies between the
bands, and this should translate into a similar loss of power spectrum
sensitivity for each band.

\begin{figure}
\begin{center}
\includegraphics[width=0.8\linewidth]{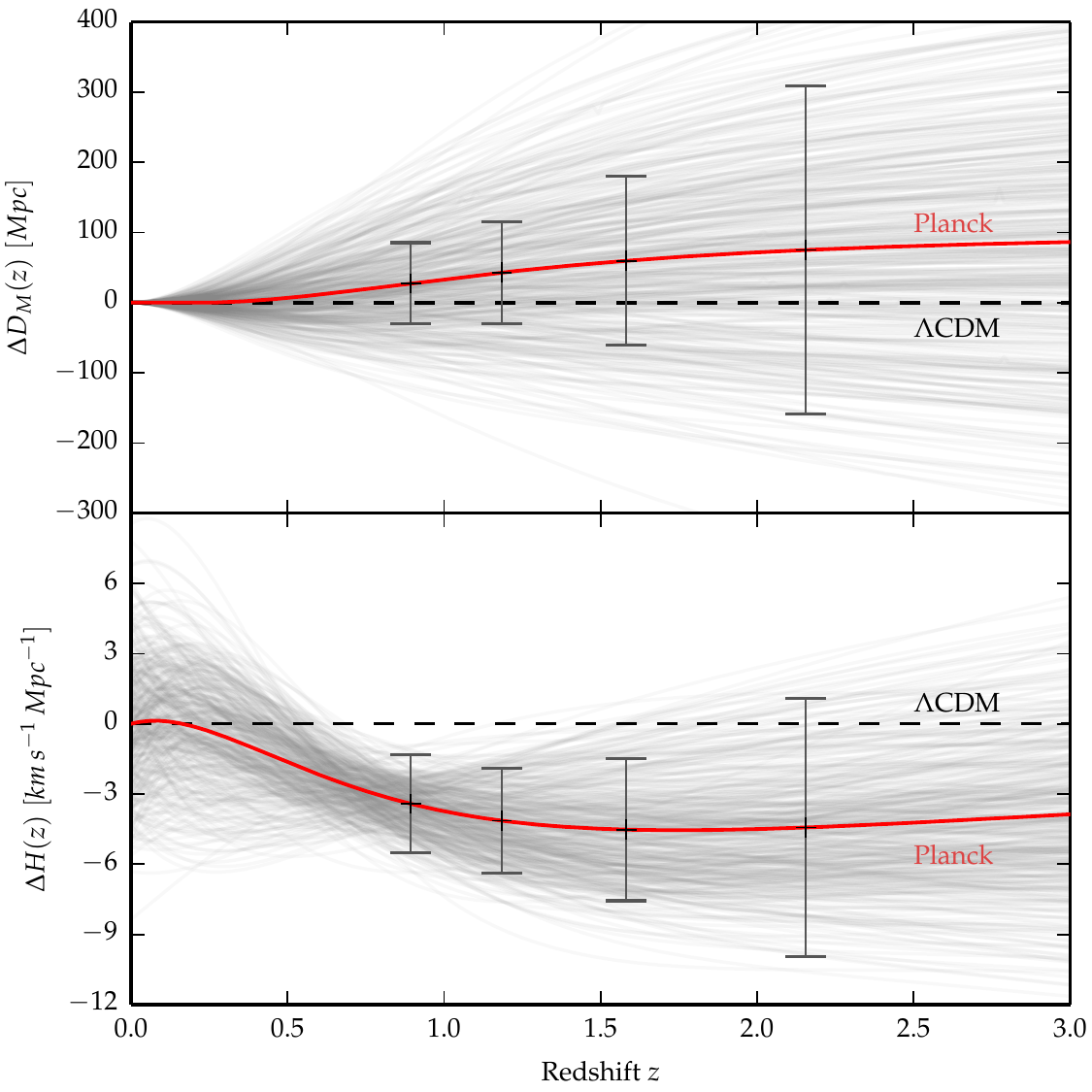}
\end{center}
\caption{Constraints on the expansion history as a function of redshift, shown
relative to a fiduical $\Lambda\mathrm{CDM}$ cosmology. The redline shows the
mean expansion history predicted from the Planck constraints on $w_0, w_a$
\cite{Planck_params} (combined with Union 2 supernovae data \cite{Union2}),
and the grey lines show a selection of histories randomly drawn from the
posterior distribution. For a medium sized cylinder experiment, the best
discrimination comes at low redshift from the \SIrange{600}{800}{\mega\hertz}
bands.}
\label{fig:DA_Hz}
\end{figure}

To constrain the dark energy equation of state, we will use the measured power
spectrum in each band to determine the apparent scale of the Baryon Acoustic
Oscillation as a function of redshift. The angular and line of sight scales
respectively constrain the transverse comoving distance $D_M(z)$ and the Hubble
parameter $H(z)$. These give two distinct probes of the expansion history as a
function of redshift. In \cref{fig:DA_Hz} we illustrate how measurements from
our example telescope could be used to improve current constraints from Planck.

\begin{figure}
\begin{center}
\includegraphics[width=0.8\linewidth]{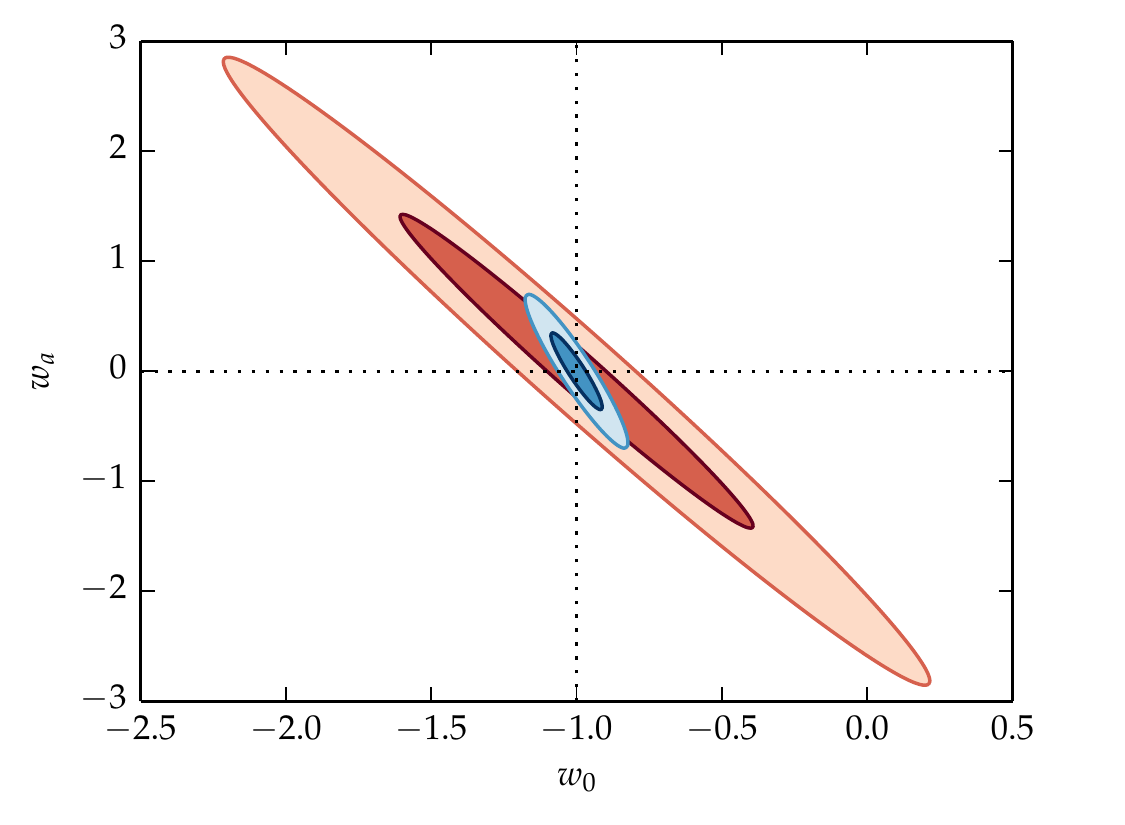}
\end{center}
\caption{Constraints on the dark energy equation of state. We show the
constraints for the example cylinder with Planck only (large, red), and with
Planck and Stage II experiments (smaller, blue). The lighter and darker
contours for each illustrate the $2 \sigma$ and $1 \sigma$ bounds
respectively.}
\label{fig:w0wa}
\end{figure}

In \cref{fig:w0wa} we show the predicted constraints on the dark energy
equation of state in the $w_0$-$w_a$ parametrisation. We describe how these
are derived from the power spectrum forecasts in \cref{app:distance}. This
gives a Figure of Merit (FoM) \cite{DETF} of $7$ for the telescope and Planck,
and $88$ if we add in Stage II experiments. This is an improvement by around
$70 \%$ from Planck and Stage II only (FoM of $53$). If there were no loss in
sensitivity due to foreground cleaning, the FoM increases to $21$ and $135$
respectively.


\section{Conclusion}
\label{sec:Conclusion}

In this paper we have improved and extended the $m$-mode formalism for
analysing observations from transit radio interferometers. In particular, we
have extended the formalism to include a complete description of polarisation
(see \cref{sec:formalism}). This allows us to characterize observations of the
real polarised sky including the effects of instrumental polarisation.
Including these effects is crucial when making wide-field multi-frequency
observations with a polarization-dependent sky response. Furthermore, by
considering the geometry of the measured data in the vector space of
observations, we have developed a simple SVD projection that not only yields a
significant data compression, but also acts as an effective filter to suppress
polarised foreground contamination (\cref{sec:svd}).

In the limit of statistically isotropic foregrounds, each $m$-mode is
independent of the others with no statistical coupling between them. Thus the
$m$-mode formalism, because it allows each mode to be treated independently,
allows for a compact and computationally efficient representation for
statistics of our data. We have exploited this to develop the KL-transform as
a technique for the removal of of astrophysical foregrounds, which otherwise
appears to be extremely challenging using other methods
(\cref{sec:kltransform}). We believe this is the first technique shown to be
effective at the removal of polarised foregrounds to below the signal level
while using a telescope model with realistic amounts of polarisation leakage
(see \cref{fig:foregroundremoval}).

Within the $m$-mode formalism we have constructed an optimal quadratic
estimator for the \tcm power spectrum that is computationally efficient and
takes into account the full statistics of the data, including the effects of
the foreground cleaning (\cref{sec:powerspectrum}). This has allowed us to
forecast the performance of a medium sized cylinder transit telescope
(\cref{sec:discussion}) --- similar in size to the CHIME pathfinder telescope
currently under construction. We show that the KL-transform is able to clean
foregrounds well into the foreground wedge, demonstrating that there is no
fundamental information loss within the region, with foreground cleaning
limiting our measurements only in a smaller band $k_\parallel
\lesssim 0.02 \ihMpc$. In fact, we find that even the removal of the polarized
foregrounds gives a minimal reduction in the expected ability to constrain the
power spectrum.

While our results are encouraging, the $m$-mode formalism does make
simplifying assumptions and the impact of these assumptions needs to be tested
when analysing real experiments. For instance, for the analysis to be
tractable, we assume the statistics of the data are stationary under rotation
of the Earth. This is expected of the \tcm signal itself, but is not expected
to be true of both for the foregrounds where the galaxy is heavily anisotropic
(though in our simulations this does not seem prevent us from suppressing
foregrounds consistent with the actual structure of the galaxy), and for the
instrumental effects where $1/f$ noise, RFI, and thermal fluctuations make the
behaviour the of the instrument time dependent. The $m$-mode formalism also
assumes perfect knowledge of the telescope, including amplifier gains, and
fully characterised beams. In \cref{sec:uncertain} we have investigated how
these uncertainties, if ignored, would lead to significant biases in the
measured power spectrum, and placed limits on how well we must know these to
faithfully recover the power spectrum. We find that random complex gain
variations can have an amplitude of up to $1 \%$ (on one minute timescales),
before they cause any significant power spectrum shifts. Similarly, using the
beam width as a simple parameterisation of our uncertainty we find that we
must know the width of the primary beam of each feed to around $0.1 \%$ to
avoid bias. These precisions are less stringent than naive expectations from
the dynamic range between the signal and foregrounds (around $10^{-5}$).
Though challenging, requiring effort and innovation, they should be achievable.

One avenue to further loosen these calibration requirements is to follow the same philosophy
we take with foreground removal and conservatively identify, and remove, the
modes which are particularly susceptible to this miscalibration. Even in the
case where we perturb the nominal beam width by an unknown number of order  $10 \%$ there is a significant fraction of the KL modes that
do not get biased appreciably. It is conceivable that through Monte-Carlo
modelling of beam uncertainties the highly corruptible KL modes could be found
and excised prior to estimating the power spectrum (at the cost of increased
error bars). Alternatively we could pursue a more targeted approach by
incorportating these instrumental uncertainties into the noise model, and
using the KL-filter to remove them. We leave investigations of these and other bias mitigating techniques to future
work.

When our feed spacing is larger than the Nyquist criterion at a
particular wavelength (for a beam stretching to the horizon this is $> \lambda
/ 2$), we cannot uniquely localise a source on the sky. This aliasing effect
causes us to form multiple images when map-making and, while not leading to
biases, gives a degradation in power spectrum errors. For the example cylinder
telescope used here, this occurs at $\nu > \SI{500}{\mega\hertz}$. While an
investigation of this effect is beyond the scope of this paper we do not
expect it is a fundamental limitation and believe that this degradation may be
alleviated by moving away from a fully uniform feed spacing.

The pipeline we have developed for performing the $m$-mode analysis described
in this paper is publically available from
\url{http://github.com/radiocosmology}. The tools created for modelling and
simulating the radio sky are available from the same location.


\noindent 

\begin{acknowledgments}

We thank the CHIME team for stimulating discussions. KS, UP, and MS are
supported in part by the Natural Sciences and Engineering Research Council
(NSERC) of Canada. The work of AS was supported by the DOE at Fermilab under
Contract No. DE-AC02-07CH11359. KS thanks Perimeter Institute for Theoretical
Physics for their hospitality. Some of the results in this paper have been
derived using the HEALPix\footnote{\url{http://healpix.sourceforge.net/}}
package \citep{HEALPIX}. Computations were performed on the GPC supercomputer
at the SciNet HPC Consortium. SciNet is funded by: the Canada Foundation for
Innovation under the auspices of Compute Canada; the Government of Ontario;
Ontario Research Fund - Research Excellence; and the University of Toronto.

\end{acknowledgments}

\appendix


\section{Noise Power Spectrum}
\label{app:noise}

The sensitivity of a a radio receiver is a well studied problem
\cite{AntennaTheory,crane:1989,wrobel:1999}. For a single feed the power
received in a frequency interval $\Delta\nu$ is simply related to the antenna
temperature $P = g^2 k_B T_a \Delta\nu$ (in the absence of noise). In our
notation the antenna temperature for a single feed is simply equal to its
auto-correlation $T_a = V_{ii}$. However, we need to extend this to the case
of the correlation of two separate antennas. Provided that the power $P
\propto \la F_i F_j^* \ra$ for both the auto-correlation $i = j$ and
cross-correlation $i \ne j$ cases, the signal observed is
\begin{equation}
P = g_i g_j^* k_B V_{ij} \Delta\nu \; .
\end{equation}
where the real and imaginary parts of $P$ contain the cosine and sine-like
correlations. The same conclusion can be reached by following through the
correlation of the induced voltage from each antenna using the effective
length. With our normalisation $\vec{l}_\text{eff}^i = l_\text{max}^i \vA_i$,
with $l_\text{max}$ the maximum length anywhere on the sky.

Beyond the astrophysical signal there are other contributions to the observed
power. This noise may come from many sources such as the ground or the
atmosphere, or the receiver system itself. For the auto-correlation of a
single feed the instantaneous noise power defines the system temperature
\begin{equation}
P = g^2 k_B T_\text{sys} \Delta\nu \; .
\end{equation}

When consider the cross-correlation between different feeds, provided the
noise at both is uncorrelated, there is no additional power observed in the
mean of the signal. However, the noise does contribute to the fluctuations
about the mean. If we average a frequency channel of width $\Delta\nu$ over a
rectangular window of time length $\tau$, we find the mean power observed is
\begin{equation}
\bar{P} = g_i g_j^* k_B \Delta\nu \lp V_{ij} + \delta_{ij} T_{\text{sys},i} \rp \; .
\end{equation}
The fluctuations in the amplitude have standard deviation
\begin{equation}
\sigma_{P} = g_i g_j^* k_B \Delta\nu \sqrt{\frac{T_{\text{sys},i}
    T_{\text{sys},j}}{\tau \Delta\nu}} \; .
\end{equation}
See \cite{crane:1989,wrobel:1999} for a detailed calculation. The fluctuations
in the real and imaginary have an equal amplitude of $\sigma_P / \sqrt{2}$. We
have assumed we are in the limit where the the system temperature dominates
the antenna temperature, $T_\text{sys} \gg T_a$.

If the noise at each feed is independent, that means that the noise between
different baseline pairs is uncorrelated. The variance that we would ascribe
to the measurement of a particular visibility $ij$ at a particular time (after
the averaging) is
\begin{equation}
\sigma_{ij}^2 = \frac{T_{\text{sys},i}(\nu)
    T_{\text{sys},j}(\nu)}{\tau \Delta\nu} \; ,
\end{equation}
To calculate the $m$-mode power spectrum of fluctuations $N_m$ we first
calculate the noise correlation function. Assuming that it is white noise, and
again using a rectangular window function the correlation function is
\begin{equation}
\zeta_{ij}(t) = \la n_{ij}(t') n_{ij}^*(t' - t) \ra = \sigma_{ij}^2 \tri{(t / \tau)} \; ,
\end{equation}
where the triangle function $\tri{(x)} = 1 - \left\lvert x \right\rvert$ for
$\left\lvert x \right\rvert < 1$. To calculate the noise power spectrum we
simply fourier transform this quantity. As we need to consider the problem in
terms of Earth rotation, we identify distinct sidereal days as independent
measurements of the sky and treat the averaged noise as periodic. Similarly we
can identify redundant baselines, as independent measurements of the same
quantity. Only the diagonal elements of the noise matrix, corresponding to the
same frequency and baseline are non-zero. The discrete  power spectrum of the
noise, defined by $\la n^m_{ij} n^{m' *}_{ij} \ra = N_{ij}^m \delta_{m m'}$, is
\begin{equation}
N_{ij}^m = \frac{T_{\text{sys},i}(\nu) T_{\text{sys},j}(\nu)}{N_\text{day} N_\text{red} t_\text{sid}\Delta\nu} \sinc^2{\left(\pi \frac{m \tau}{t_\text{sid}}\right)} \; ,
\end{equation}
where $N_\text{day}$ is the number of sidereal days that have been
observed. Usually we would want the integration length to be
smaller than any angular scale we are interested in, in this limit $m \tau \ll
t_\text{sid}$, and the $\sinc$ factor is $\sim 1$.


\section{\KLfull Transform}

\label{app:snmodes}

Let us write our measurement as a vector $\vx$, where the each dimension
corresponds to a measured degree of freedom. We can write $\vx$ as
\begin{equation}
\vx = \vs + \vn
\end{equation}
where $\vs$ and $\vn$ are respectively the signal we are interested in and some
generalised form of noise (in the case of \tcm this may include the
foregrounds). These components have covariance matrices
\begin{equation}
\la \vs \vs^\hconj\ \ra = \mS, \qquad \la \vn \vn^\hconj \ra = \mN \; .
\end{equation}
We are free to transform the measurement vector as we wish, $\vx' = \mR\vx$,
provided we are careful to update all the statistics we make use of. In our
case we are interested in the two point statistics and so it is sufficient to
transform the covariance matrix $\mat{X}' = \la (\mR \vx) (\mR \vx)^\hconj \ra
= \mR \mat{X}\mR^\hconj$. The Karhunen-Loeve (KL) transform takes advantage of
this to produce simultaneous eigenmodes of the signal and noise covariances.

We start by making the eigendecomposition of the noise matrix
\begin{equation}
\mN = \mR_1^\hconj \mN' \mR_1
\end{equation}
where $\mR_1$ is the unitary matrix of eigenvectors (stacked row by row), and
$\mN'$ is the diagonal matrix of eigenvalues. Using this we can transform the
data vector $\vx' = \mR_1 \vx$, which produces a new signal covariance
\begin{equation}
\mS' = \la \vs' \vs'^\hconj \ra = \la (\mR_1 \vs)  (\mR_1 \vs)^\hconj
\ra = \mR_1 \mS \mR_1^\hconj
\end{equation}
and reduces the noise matrix to $\mN'$. As the new noise matrix consists solely
of positive diagonal elements $(\mN')_{ii} = \lambda^N_i$, a further
transformation $\vx'' = \mR_2 \vx'$, where $\mR_2 = \mN'^{-\frac{1}{2}}$,
reduces the noise matrix to the identity $\mN'' = \mat{I}$. The signal matrix
is transformed to
\begin{equation}
\mS'' = \mR_2 \mR_1 \mS \mR_1^\hconj \mR_2^\hconj \; .
\end{equation}

Applying any unitary transformation to the data will leave the noise covariance
as the identity. We use this freedom to diagonalise the signal covariance by
eigendecomposition $\mS'' = \mR_3^\hconj \mLambda \mR_3$, leaving the total
transformation on the data as
\begin{equation}
\vx \rightarrow \vxt = \mR_3 \mR_2 \mR_1 \vx \; .
\end{equation}
Overall this has changed the covariance matrices to
\begin{align}
\mS & \rightarrow \mLambda \; ,\\
\mN & \rightarrow \mI \; \; .
\end{align}
By making this transformation we have simultaneously diagonalised the
correlations of both the signal and the noise, mapping the latter to the
identity matrix. In particular, the elements of $\mLambda$ give the signal to
noise ratio of each mode. With no hidden correlations this basis allows us to
cleanly filter data by simply throwing away modes with signal to noise ratio
below some threshold. This is equivalent to zeroing the corresponding elements
of $\vxt$.

Rather than explicitly constructing the three transformations, it is
mathematically equivalent to find the solutions to the generalised eigenvalue
problem
\begin{equation}
\mS \vx = \lambda \mN \vx \; ,
\end{equation}
with the eigenvectors forming the transformation matrix, and the eigenvalues
giving the elements of the signal covariance $\mLambda$. This approach is
simpler and computationally more efficient.


\section{Statistical Models}
\label{app:models}

As discussed in \cref{sec:kltransform} to use the \KLfull transform to perform
foreground cleaning we require models of the two-point statistics of both the
\tcm signal and the foreground contaminants. For computational efficiency
these models must be isotropic and so we only need to specify the angular power
spectrum
\begin{equation}
C_l^{XY}(\nu, \nu') = \la a_{lm}^X(\nu) a_{lm}^{Y*}(\nu') \ra \; ,
\end{equation}
for all the pairs of the four polarisation components $X, Y \in \left\{T, E,
B, V\right\}$.

\subsection{Astrophysical Foregrounds}

Our foreground models are based on \cite{SantosCoorayKnox}. However we only
include the dominant two components, the galactic synchrotron emission and
extragalactic point sources. In both cases the angular power spectrum is of the
form
\begin{multline}
\label{eq:aps_sck}
C_l(\nu, \nu') = A \lp \frac{l}{100}\rp^{-\alpha} \! \lp \frac{\nu \nu'}{\nu_0^2} \rp^{-\beta} \! e^{ - \frac{1}{2 \xi^2_l} \ln^2{(\nu / \nu')}} \, .
\end{multline}

The original models were calibrated for forecasting observations of Epoch of
Reionisation. In \cite{Shaw2013} we recalibrated them for the high frequency,
all sky observations we are concerned with in this paper. However for this
work we also need to specify the correlations of the polarised parts of the
foregrounds. We assume that the dominant source of polarised emission is our
own galaxy (ignoring the polarisation of point sources) and model the
polarised emission as being a statistical fraction $f_\text{pol}$ of the
unpolarised emission
\begin{equation}
C_l^{EE}(\nu, \nu') = C_l^{BB}(\nu, \nu') = f_\text{pol}^2 C_l^{TT}(\nu, \nu') \; .
\end{equation}
In addition we assume that the polarised emission is uncorrelated such that
$C_l^{TE} = C_l^{TB} = C_l^{EB} = 0$, and that there is no circular
polarisation from the galaxy $C_l^{VV} = 0$. Our fiducial polarisation
fraction is $f_\text{pol} = 0.5$. We list the parameters for these models in
\cref{tab:modelparams}.

\begin{table}

\caption{Parameters for our foreground power spectrum model given in
\eqref{eq:aps_sck}. These are based on the models of \cite{SantosCoorayKnox},
adapted to the intensity mapping regime in \cite{Shaw2013}.}

\label{tab:modelparams}
\begin{ruledtabular}
\begin{tabular}{l@{\hspace{0.04\linewidth}}l@{\hspace{0.04\linewidth}}l@{\hspace{0.04\linewidth}}l@{\hspace{0.04\linewidth}}l@{\hspace{0.04\linewidth}}l}
Component       & Polarisation & A (\si{\kelvin\squared})  & $\alpha$  & $\beta$   & $\zeta$   \\
\hline
Galaxy          & TT & \num{6.6e-3}              & \num{2.80}& \num{2.8} & \num{4.0} \\
                & EE, BB & \num{1.65e-3}         & \num{2.80}& \num{2.8} & \num{4.0} \\
Point Sources   & TT & \num{3.55e-4}             & \num{2.10}& \num{1.1} & \num{1.0} \\
\end{tabular}
\end{ruledtabular}

\end{table}

\subsection{\tcmt Signal}

On large scales the \tcm brightness temperature is a biased tracer of the matter density field \cite{Masui2013} with a power spectrum $P_{T_b}$ given by
\begin{equation}
\label{eq:psreal}
P_{T_b}(\vk; z, z') = \bar{T}_b(z) \bar{T}_b(z') \lp b + f \mu^2 \rp^2 P_m(k; z, z')
\end{equation}
where $b$ is the bias and $P_m(k; z, z') = P(k) D_+(z) D_+(z')$ is the
real-space matter power spectrum. The evolution of the perturbations is given
by the growth factor $D_+(z)$ normalised such that $D_+(0) = 1$, with the
growth rate $f = d\ln{D_+} / d\ln{a}$ (that is the logarithmic derivative of
the growth factor $D_+$). The mean brightness temperature is assumed to take
the form
\begin{multline}
\label{eq:mean_temp}
\bar{T}_b(z) =  0.1 \lp \frac{\Omega_\text{HI}}{0.33\times 10^{-4}}\rp \\ \times\lp \frac{\Omega_m + (1+z)^{-3}\Omega_\Lambda}{0.29} \rp^{-1/2} \lp \frac{1+z}{2.5}\rp^{1/2} \si{\milli\kelvin}
\end{multline}
given in \cite{Chang2008}. In \cite{Switzer2013} they determine the degenerate
product $\Omega_\text{HI} b = 0.62 \times 10^{-3}$, which we use in this work.
As the redshift distortions break the $\Omega_\text{HI} b$ degeneracy we fix
$b = 1$.

For use in our foreground filter, we require the angular power spectrum of the
\tcm brightness temperature \citep{Lewis2007,Datta2007}. This can be
calculated from the real-space power spectrum \eqref{eq:psreal}, but is
compuationally difficult, generally requiring double-integration over highly
oscillatory functions for each $\nu$, $\nu'$ pair. To speed this up we use the
flat-sky approximation from
\cite{Datta2007}
\begin{equation}
\label{eq:cl_flatsky}
C_l(z, z') = \frac{1}{\pi \chi \chi'} \int_0^\infty \!\! dk_\parallel \cos{\lp k_\parallel \Delta\chi\rp} P_{T_b}(\vk; z, z')
\end{equation}
where $\chi$ and $\chi'$ are the comoving distances to redshift $z$ and $z'$
and their difference is denoted by $\Delta\chi = \chi - \chi'$. The wavevector
$\vk$ has components $k_\parallel$ and $l / \bar{\chi}$ in the directions
parallel and perpendicular to the line of sight ($\bar{\chi}$ is the mean of
$\chi$ and $\chi'$). This approximation is accurate to the 1\% level for $l >
10$ \citep{Datta2007}.

We use this method not only for calculating the signal covariance function,
but also the band functions required for the power spectrum. To determine each
$\mC_a$ we simply apply \eqref{eq:psreal} and \eqref{eq:cl_flatsky}, with
$P_m(\vk) = P_a(\vk)$.


\section{Simulating All-sky Radio Emission}
\label{app:simulatedmaps}

Testing of the $m$-mode formalism, and the foreground cleaning with the
\KLfull transform requires the use of synthetic sky maps. For it to be
realistic these simulated maps must capture the essential properties of the
\tcm signal and foreground components. In this Section we briefly describe how
these simulations are generated.

\subsection{\tcmt Signal}

Assuming the cosmological \tcm emission is Gaussian on the scales of interest,
the angular power spectrum given in the previous section
(\eqref{eq:cl_flatsky}) completely specifies its fluctuations. Maps of the sky
can be generated by drawing Gaussian realisations of the power spectrum, using
Cholesky decomposition to produce the correct frequency correlation structure,
and then adding in the mean temperature given by \eqref{eq:mean_temp}.

\subsection{Extra-Galactic Point Sources}

We construct our point source simulations from three components: a population
of real bright point sources ($S > \SI{10}{\jansky}$ at
\SI{151}{\mega\hertz}); a synthetic population of dimmer sources down to
\SI{0.1}{\jansky} at \SI{151}{\mega\hertz}; and an unresolved background of
dimmer sources ($S < \SI{0.1}{\jansky}$) modelled as a Gaussian random field.
This last component dramatically reduces the number of sources we must
directly generate.

The unresolved background is generated by drawing a Gaussian realisation from
the point source model detailed in \cref{tab:modelparams}. The random source
catalogue is constructed by drawing from the point source distribution of
\cite{DiMatteo2002} and scattering the sources randomly over the sky. The
intrinsic polarisation of each point source is determined by
\begin{equation}
Q(\nu) + i U(\nu) = p I(\nu) \,
\end{equation}
where the polarisation fraction $p$ is a complex Gaussian random variable with
standard deviation $\sigma_p$. This standard deviation is equal to the average
polarisation fraction of sources in the catalogue, we set $\sigma_p = 5\%$.

The population of real bright point sources is generated by matching VLSS at
\SI{74}{\mega\hertz} \cite{VLSS} against NVSS at \SI{1.4}{\giga\hertz}
\cite{NVSS}. We only include sources interpolated to be brighter than
\SI{10}{\jansky} at \SI{151}{\mega\hertz}. Each source is assigned the
polarisation as measured by NVSS, and is extrapolated to other frequencies. In
this work we have also assumed that the six sources above \SI{100}{\jansky}
(at \SI{600}{\mega\hertz}) have been removed from the timestream to high
accuracy.

The polarisation of an extra-galactic source is Faraday rotated as it passes
through the magnetised interstellar medium in our galaxy, generating
oscillatory frequency structure in the polarisation. To apply this, we use the
Faraday depth map of \cite{Oppermann2012} to rotate the polarisation angle of
our background sources.

\subsection{Galactic Synchrotron Intensity}

In this work we continue to use the prescription developed in a previous paper
\cite{Shaw2013} to generate constrained simulations of the total intensity of
synchrotron emission from our galaxy. These maps are formed from two distinct
components:

\begin{itemize}

\item A large scale base map produced by extrapolating the Haslam
map\footnote{We use the map from the Legacy Archive for Microwave Background
Data Analysis (LAMBDA), which has been processed to remove bright point
sources and striping. See
\url{http://lambda.gsfc.nasa.gov/product/foreground/haslam_408.cfm}} with a
spectral index map from \cite{MD2008}.

\item A randomly generated map that adds in fluctuations in frequency and on
small angular scales. This is constrained to be zero on the scales constrained
by the Haslam map, and is designed to smoothly extrapolate the angular
fluctuations of the Haslam map to smaller scales, and reproduce the
anisotropic fluctuations on small scale power across the sky
\cite{LaPorta2008}.

\end{itemize}
The procedure for generating these two components is described in detail in
\cite{Shaw2013}, with the only change being the spectral index map used.

\subsection{Galactic Synchrotron Polarisation}

\renewcommand{\vr}{\ensuremath{\vec{r}}}

To test our foreground removal and analysis we need to be able to create
simulated multi-frequency maps of our galaxy, and in particular its
polarisation structure. As the observed radiation has been omitted across a
range of Faraday depths, unlike extra-galactic sources, this is challenging.
One approach is to make use of the increasingly sophisticated models of the
galactic magnetic field structure \cite{Jansson2012}, and electron
distribution \cite{NE2001}, to create realistic large scale simulations of the
polarisation structure \cite{Hammurabi}. However, we instead appeal to the
ideas of Faraday Rotation Measure Synthesis \cite{Brentjens2005} to rapidly
create simulations that capture the important effects.

Rotation measure synthesis attempts to link the wavelength dependent
polarisation rotation to the structure along the line of sight. Polarised
radiation emitted at a distance $r$ from us is Faraday rotated by an amount
$\phi \lambda^2$ before it reaches us, where the Faraday depth
\begin{equation}
\phi(\vr) = \int_0^r n_e(\vr') \vec{B}(\vr') \cdot d\vr' \; .
\end{equation}
The key idea in Faraday Rotation Measure Synthesis is to not directly probe
the physical structure of emission, but to probe the structure as a function
of Faraday depth. In this case we can just think of the observed polarised
emission in a given direction $P(\vnhat, \lambda^2)$ as being the summation of
the emission at all Faraday depths $F(\vnhat, \phi, \lambda^2)$, rotated by
the correct wavelength dependent amount
\begin{equation}
\label{eq:rm_synthesis}
P(\vnhat, \lambda^2) = \int F(\vnhat, \phi, \lambda^2) e^{2 i \phi \lambda^2} d\phi \; .
\end{equation}
In \cite{Brentjens2005} the idea was to use multi-wavelength observations to
invert this Fourier relation, and constrain the structure of $F(\vnhat, \phi,
\lambda^2)$. However, we will attempt to use well motivated assumptions about
the emission in Faraday-space to construct simulations of polarised skies. As
in \cite{Brentjens2005} we presume that the Faraday space emission $F(\vnhat,
\phi,\lambda^2)$ is separable in its spectral dependence, such that $
F(\vnhat, \phi, \lambda^2)  = f(\vnhat, \phi) s(\vnhat, \lambda^2)$. This
flattens the spectrum of $f$ so that we can still use the Faraday synthesis
formalism for it. We take the spectral function $s(\vnhat, \lambda^2)$ from
the unpolarised emission.

We start with a simple model of the emission from the galaxy, assuming that
along any line of sight the emission comes from many independent synchrotron
regions each of fixed brightness $\Delta{T}$. In a direction with total
brightness temperature $T$, there are $N = T / \Delta{T}$ such regions. We
assume that the emitting regions are scattered across a range in Faraday depth.
With no reason to favour positive or negative Faraday depths, we assume this
distribution is zero mean. From observations of extragalactic point sources we
know the Faraday depth to the edge of our galaxy \cite{Oppermann2012} and this
gives us a measure of the range of Faraday depths within the galaxy. Combining
these properties the distribution is modelled as a zero-mean Gaussian with a
width $\sigma_\phi(\vnhat)$ which is determined from the Faraday rotation data.
We determine $\sigma_\phi(\vnhat)$ by taking the Faraday depth map of
\cite{Oppermann2012}, taking its absolute value, and smoothing with a FWHM of
\SI{10}{\degree}. Each of these regions has a small width in $\phi$ over which
its polarisation is coherent. We call this coherence length $\xi_\phi$, and
note that it determines size of structures in Faraday space


To determine the polarisation structure we start by calculating the number of
emitting regions within a range $\phi$ to $\phi + \Delta{\phi}$. This is given
by
\begin{equation}
\Delta{N} = \frac{N}{(2 \pi \sigma_\phi^2 )^{1/2}} e^{-\frac{1}{2} \lp\frac{\phi}{\sigma_\phi}\rp^2} \Delta{\phi}
\end{equation}
As each region is independent we assume they have a randomly distributed
complex polarisation, drawn from a Gaussian distribution with variance
$(\alpha_p \Delta{T})^2$, where $\alpha_p$ is the polarisation fraction.
Within this range in Faraday depth the polarisations add up like a random
walk, giving the expected total root-mean-square polarisation as
\begin{multline}
\alpha_p \Delta{T} \Delta{N}^{1/2} = (8 \pi)^{1/4} \alpha_p \Delta_T \\ \times \ls \frac{1}{(4 \pi \sigma_\phi^2 )^{1/2}} e^{-\frac{1}{4} \lp\frac{\phi}{\sigma_\phi}\rp^2} \rs \lp \frac{N \sigma_\phi}{\Delta{\phi}} \rp^{\frac{1}{2}}  \Delta{\phi} \; .
\end{multline}
This gives the expected magnitude of the emission at each position in Faraday
space, showing that even in Faraday space we see depolarisation because of the
incoheret combination of multiple Faraday sources at a single depth.

This suggests we model the emission as two factors
\begin{equation}
f(\vnhat, \phi) = w(\vnhat, \phi) c(\vnhat, \phi) \; .
\end{equation}
The first $w(\vnhat, \phi)$ is a positive envelope function which defines the
region, and amplitude of emission in Faraday depth.
\begin{equation}
w(\vnhat, \phi) \propto \frac{A}{\sqrt{4\pi \sigma_\phi^2}} e^{-\frac{1}{4} \lp\frac{\phi}{\sigma_\phi}\rp^2} (T \sigma_\phi)^{1/2}
\end{equation}
The second $c(\vnhat, \phi)$ is a random field that gives fluctuations in the
complex polarisation as a function of Faraday depth, this should be highly
correlated on scales $\Delta\phi \ll \xi_\phi$, and uncorrelated on scales
$\Delta\phi \gg \xi_\phi$. We model this as a Gaussian random field drawn with
an angular power spectrum
\begin{equation}
\label{eq:c_corr}
C_l(\phi, \phi') \propto \lp \frac{l}{100} \rp^{-\alpha} \exp{\lp -\frac{(\phi - \phi')^2}{2 \zeta^2}\rp} \; .
\end{equation}
The angular dependence is chose to match that of the total intensity model
\eqref{eq:aps_sck}.

The normalisation of these functions is degenerate with the value of
$\Delta{T}$. We fix the combination by considering what happens at high
frequency observations where Faraday rotation is much less important. In this
limit the polarisation fraction is determined by the incoherent addition of
the polarisation of the emitting regions and is $\sim \alpha_p (\Delta{T} /
T)^{1/2}$. We choose $\alpha_p = 2/3$ which is the intrinsic polarisation of
synchrotron with a spectral index of the electron energy distribution $\gamma
= 5/3$. Polarised maps from the WMAP satellite at \SI{23}{\giga\hertz}
\cite{Kogut2007} indicate that the galaxy is 20\% polarised at high latitudes,
we use this fact to determine the overall normalisation.

The only remaining degree of freedom is the the correlation length of the
emitting regions if Faraday space, $\xi_\phi$. The size of an emitting region
in Faraday space will grow towards the galactic centre because of the
increased magnetic field strengths. We construct a crude model
\begin{equation}
\xi_\phi = \min{\lp \sigma_\phi / 20,\: \SI{3}{\radian\per\square\metre}\rp} \; ,
\end{equation}
chosen to visually reproduce the amount of depolarisation seen in
\SI{1.4}{\giga\hertz} polarisation maps
\cite{Wolleben2006,Testori2008,Reich2009}.

This gives all the necessary ingredients to draw a realisation of $f(\vnhat,
\phi)$ which we can Fourier transform, and scale by the spectral function
$s(\vnhat, \lambda^2)$ to produce the polarised emission using
\eqref{eq:rm_synthesis}. All these operations are performed on a regular grid
in $\lambda^2$, which is extended beyond the desired frequency range to negate
edge effects. The resulting series of maps are then interpolated onto the
required frequency slices.

\begin{figure}

\begin{center}
\textbf{Polarisation Fraction}\\
\includegraphics[width=\linewidth]{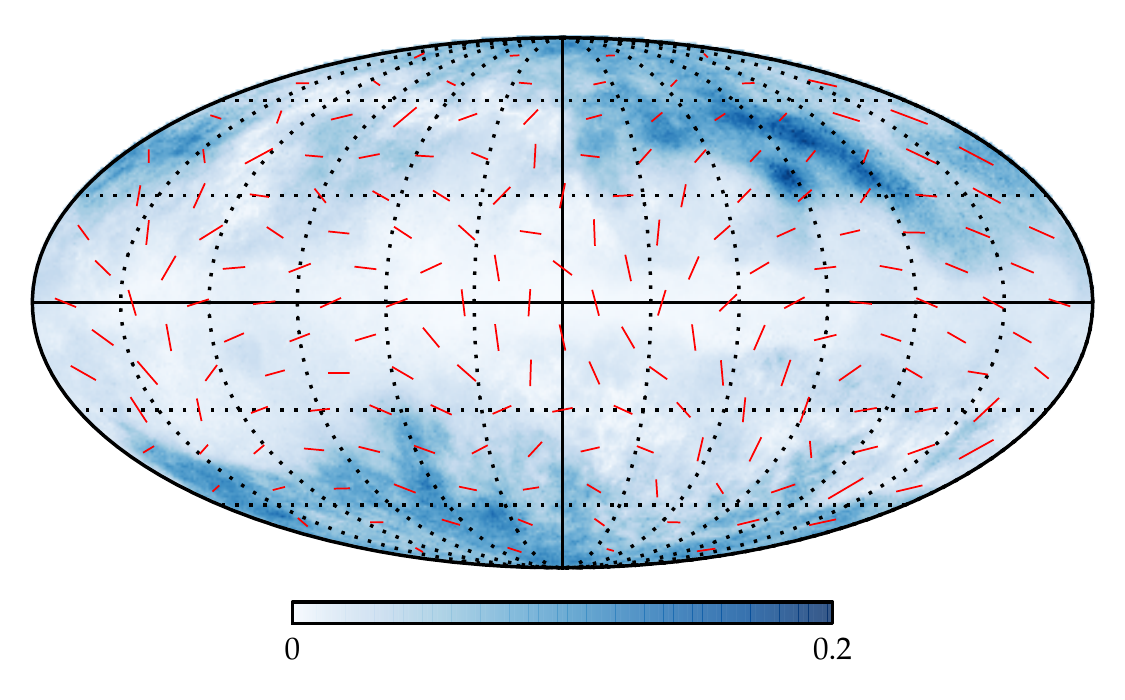}\\[0.5cm]
\textbf{Correlation length}\\
\includegraphics[width=\linewidth]{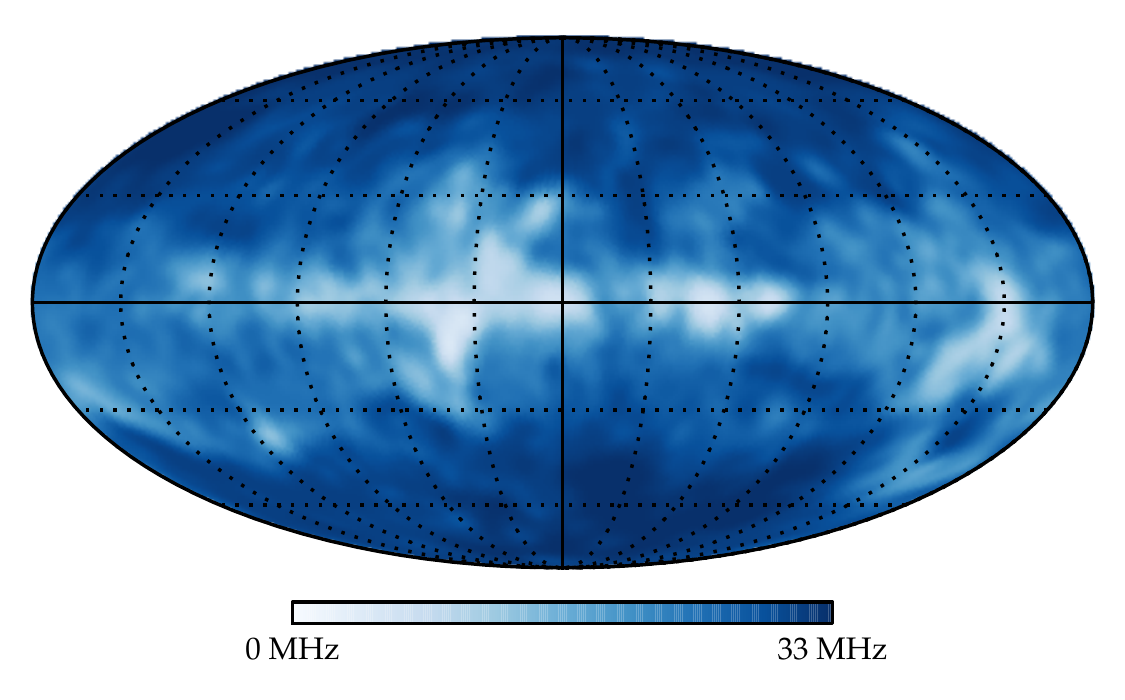}
\end{center}

\caption{The top figure shows the polarisation direction and fraction of the
\SI{600}{\mega\hertz} slice of a polarised simulation of the galaxy. This
clearly demonstrates the effect of Faraday depolarisation towards the galactic
centre. The lower plot show the effective correlation length as measured
across the sky, smoothed on \SI{10}{\degree} scales.}

\label{fig:polsim}

\end{figure}

In \cref{fig:polsim} we show the polarisation fraction, and frequency
correlation length derived from a simulation between
\SIrange{400}{600}{\mega\hertz}. Though the model we have constructed here is
crude, and based unrealistic assumptions about the galactic emission it
exhibits the properties we would expect from the real galactic emission:

\begin{itemize}

\item Emission is from a range of Faraday depths, rather than a single screen.

\item In the galactic plane there is substantial depolarisation, but at high
latitudes the polarisation fraction is around that at \SI{23}{\giga\hertz}.

\item Frequency decorrelation on lengths that we would predict from the
Faraday rotation over the galaxy, going to near zero in the galactic centre
where the emission goes up to large Faraday depths.

\end{itemize}

\section{Distance Measurements}
\label{app:distance}

By extracting the Baryon Acoustic Oscillation (BAO) signal from the \tcm power
spectrum, measurements of the Hubble rate $H(z)$ and transverse comoving
distance $D_M(z)$ can be made. The BAO manifests itself as a preferred
separation in the two-point correlation function at distances $s_{\perp}$
perpendicular to the line of sight and $s_{\parallel}$ parallel to the line of
sight. The fractional errors on $s_{\perp}$ and $s_{\parallel}$ are equivalent
to the fractional errors on the combinations $s/D_M$ and $s H$, respectively,
where $s$ is the comoving sound horizon at the drag epoch. Thus, if $s$ is
well-known (for example from observations of the Cosmic Microwave Background)
then measurements of $s_{\perp}$ and $s_{\parallel}$ put observational
constraints on $D_M$ and $H$.

To project uncertainties in the power spectrum onto $D_M$ and $H$, we first
transform the Fisher matrix for the power spectrum $\mF$ into the Fisher matrix
$\mF_s$ for the parameters $\theta_{s}=(\ln s_{\perp}^{-1}, \ln s_{\parallel})$
via the Jacobian $(J_{s})_{ij} = \partial P(\textbf{k}_i) / \partial
(\theta_{s})_j$, where the Fisher matrices are related by $\mF_{s} = \mJ_{s}^T
\mF \mJ_{s}$. $\mJ_{s}$ is evaluated using a fiducial cosmological model.

To calculate $\mJ_{s}$, we follow \cite{Seo2007}, where the effect of the
baryons on the power spectrum is modelled by an additive term $P_{\rm{b}}$ to
the otherwise `wiggles-free' power spectrum that is approximated as
\begin{multline}
P_{\rm{b}}(\textbf{k}) = \sqrt{8\pi^2} A_0 P_{0.2} \sinc(x) \\ \times \exp{\ls -(k/k_{s})^{1.4} -(k\Sigma_{\rm{nl}})^2/2 \rs}
\label{Pbao}
\end{multline}
where $x=\sqrt{(k_{\perp}s_{\perp})^2 + (k_{\parallel}s_{\parallel})^2}$,
$k_{s}$ is the Silk scale, and  $P_{0.2}$ is the linear power spectrum
evaluated at $k=0.2 \, h \rm{Mpc}^{-1}$. In this expression, $A_0$ is a
normalization constant, taken to be $A_0=0.5817$. $\Sigma_{\rm{nl}}$ is the nonlinear dampening scale given by
\begin{equation}
	\Sigma_{\rm{nl}}^2 = (1-\mu^2)\Sigma_{\perp}^2 + \mu^2\Sigma_{\parallel}^2
\end{equation}
where $\Sigma_{\parallel}=\Sigma_{\perp}(1+f)$, $\Sigma_{\perp}= \Sigma_0 (G(z)/G(0))$, $G$ is the growth function, and $f$ is the growth rate. We follow \cite{Seo2010} and assume that we may partially reconstruct parts of the BAO signal degraded by nonlinear effects for modes with high signal to noise and set the effective nonlinear dampening scale to be $\Sigma_0 = 4.70 (\sigma_8/0.9) h^{-1} \rm{Mpc}$. By differentiating \cref{Pbao} with respect to the variables $\theta_{s}$, one can form $\mJ_{s}$ and subsequently evaluate $\mF_{s}$. Note that $\mF_{s}$ is equivalent to the Fisher matrix for the variables $\theta_{\rm{d}} =
( \ln (D_M(z)/s), \ln (sH(z)) )$.

The Fisher matrix $\mF_{s}$ can be transformed again into the Fisher matrix
$\mF_{\text{DE}}$ for the cosmological parameters $\theta_{\text{DE}} = ( w_0,
w_a, \Omega_{\Lambda}, \Omega_k, \omega_m, \omega_b )$ by use of the Jacobian
$(\mJ_{\text{DE}})_{ij}= \partial (\theta_{\rm{d}})_i / \partial
(\theta_{\text{DE}})_j $, which as before is evaluated using a fiducial
cosmological model. In $\theta_{\text{DE}}$, the equation of state $w$ has been
parameterized as
\begin{equation}
w(z) = w_0 + w_a \frac{z}{1+z}
\end{equation}
Note that the $\omega_b$ dependence in $\mF_\text{DE}$ comes from the comoving
sound horizon $s$, present in both terms of $\theta_{\rm{d}}$, which is
dependent on the baryon to photon ratio $R_b = 3\rho_b/4\rho_{\gamma}$. The
Fisher matrix for the dark energy parameters is then formed as $\mF_{\text{DE}} =
\mJ_{\text{DE}}^{\rm{T}} \mF_{s} \mJ_{\text{DE}}$. Constraint contours in the
$w_0-w_a$ plane can be found by marginalizing over the other variables in
$\theta_{\text{DE}}$, which in this case amounts to inverting
$\mF_{\text{DE}}$ to get the covariance matrix, removing the rows and columns
corresponding to the marginalized variables, and inverting once more to
recover the marginalized Fisher matrix for $(w_0,w_a)$.






\bibliography{cyl,telescopes}

\end{document}